\documentclass[usenatbib,usegraphicx]{mn2e}
\def\msun{{\rm M}_{\odot}}
\def\rsun{{\rm R}_{\odot}}

\begin{document}

\title[Rotation of low-mass stars in M34]{The {\em Monitor} project:
    Rotation of low-mass stars in the open cluster M34}
\author[J.~M.~Irwin et al.]{Jonathan~Irwin$^{1}$, Suzanne~Aigrain$^{1}$,
  Simon~Hodgkin$^{1}$, Mike~Irwin$^{1}$,
\newauthor
Jerome~Bouvier$^{2}$, Cathie~Clarke$^{1}$, Leslie~Hebb$^{3}$,
    Estelle~Moraux$^{2}$ \\
$^{1}$Institute of Astronomy, University of Cambridge, Madingley Road,
  Cambridge, CB3 0HA, United Kingdom \\
$^{2}$Laboratoire d'Astrophysique, Observatoire de Grenoble, BP 53,
  F-38041 Grenoble C\'{e}dex 9, France \\
$^{3}$School of Physics and Astronomy, University of St Andrews,
  North Haugh, St Andrews, KY16 9SS, Scotland}
\date{Accepted 2006 May 05.  Received 2006 May 05; in original form 2006 April 18}

\maketitle

\begin{abstract}
We report on the results of a $V$ and $i$-band time-series photometric
survey of M34 (NGC 1039) using the Wide Field Camera (WFC) on the
Isaac Newton Telescope (INT), achieving better than $1\%$ precision
per data point for $13 \la i \la 17$.  Candidate cluster members were
selected from a $V$ vs $V-I$ colour-magnitude diagram over $14 < V <
24$ ($0.12 \la M/\msun \la 1.0$), finding $714$ candidates, of which
we expect $\sim 400$ to be real cluster members (taking into account
contamination from the field).  The mass function was computed, and
found to be consistent with a log-normal distribution in ${\rm
  dN/dlogM}$. Searching for periodic variable objects in the candidate
members gave $105$ detections over the mass range $0.25 < M/\msun <
1.0$.  The distribution of rotation periods for $0.4 < M/\msun < 1.0$
was found to peak at $\sim 7\ {\rm days}$, with a tail of fast
rotators down to periods of $\sim 0.8\ {\rm days}$.  For $0.25 <
M/\msun < 0.4$ we found a peak at short periods, with a lack of slow
rotators (eg. $P \ga 5\ {\rm days}$), consistent with the work of
other authors (eg. \citealt{se2004}) at very low masses.   Our results
are interpreted in the context of previous work, finding that we
reproduce the same general features in the rotational period
distributions.  A number of rapid rotators were found with velocities
$\sim$ a factor of two lower than in the Pleiades, consistent with
models of angular momentum evolution assuming solid body rotation
without needing to invoke core-envelope decoupling.
\end{abstract}

\section{Introduction}
\label{intro_section}

M34 (NGC 1039) is a relatively well-studied intermediate-age open
cluster ($\sim 200\ {\rm Myr}$, \citealt{m93}; \citealt{jp96}), at
a distance of $\sim 470\ {\rm pc}$, distance modulus 
$(M-m)_0 = 8.38$ \citep{jp96}, with low reddening:
$E(B-V) = 0.07\ {\rm mag}$ \citep{ccp79}.  The cluster is
located at $\alpha = {\rm 2^h42^m06^s}$, $\delta = {\rm +42^\circ46'}$
(J2000 coordinates from SIMBAD), corresponding to galactic coordinates
$l = 143.7^\circ$, $b = -15.6^\circ$.

Clusters in the age range between $\alpha$ Persei ($\sim 80\ {\rm
  Myr}$, \citealt{s99}) and the Hyades ($\sim 625\ {\rm Myr}$,
\citealt{p98}) are important for constraining stellar evolution
models.  A $1 \msun$ star reaches the zero age main sequence (ZAMS)
at $\sim 30\ {\rm Myr}$, with progressively later spectral types
reaching the ZAMS at later times.  M34 is useful in this context owing
to both its age and relative proximity, where we can study solar
mass stars after evolving for a short time on the main sequence, and
lower mass stars as they finish pre main sequence evolution.  We
estimate that $0.5\ \msun$ stars reach the ZAMS at approximately the
age of M34, using the models of \citet{bcah98}.

Membership surveys of M34 using proper motions (\citealt{is93},
\citealt{jp96}) have been carried out using photographic
plates down to $V \sim 16.2$ ($M \sim 0.7\ \msun$), finding a mean
cluster proper motion of $\sim 20\ {\rm mas\ yr^{-1}}$.  Studies of M34
members in the literature include spectroscopic abundance measurements
for G and K dwarfs, finding a mean cluster metallicity $[{\rm Fe/H}] =
+0.07 \pm 0.04$ \citep{s2003}.  Rotation studies have used both
spectroscopy \citep{sjf}, covering the high- to intermediate-mass end
of the M34 cluster members, down to G spectral types, and photometry
\citep{b2003}.  Finally, a ROSAT soft X-ray survey \citep{s2000},
detected $32$ sources.

The existing surveys in M34 have concentrated on the high-mass end of
the cluster sequence (typically K or earlier spectral types).  Later
spectral types, in particular M, down to the substellar regime, are
not well-studied.  A catalogue of low-mass M34 members could be used
to perform both photometric (eg. surveys for rotation and eclipses)
and spectroscopic (eg. spectral typing and metal abundance) studies of
the K and M dwarf populations.

\subsection{Evolution of stellar angular momentum}
\label{amevol_section}

The early angular momentum distribution of young stars is commonly
explained as resulting from an initial intrinsic angular momentum
distribution consisting of relatively mild rotators (eg. with
rotational periods of $\sim 8\ {\rm days}$ for typical very young
classical T Tauri stars such as the population of slow rotators in the
ONC, \citealt{h2002}), modified by subsequent spin-up of the star
during contraction on the PMS.  Observations indicate that a fraction
of the stars spin slower than expected due to contraction, suggesting
that they are prevented from spinning up.  The most popular
explanation for this is disc locking (eg. \citealt{k91},
\citealt*{bfa97}, \citealt{cc95}).  In the T-Tauri phase, stars are
still surrounded by accretion discs, so a star could retain a constant
rotation rate during contraction due to angular momentum transport via
magnetic interaction of the star and disc.  The stars are thus
prevented from spinning up until the accretion disc dissolves.  Recent
evidence in favour of this hypothesis has been obtained from Spitzer
studies, eg. \citet{rebull06}.

Very young clusters such as the Orion Nebula Cluster (ONC, age $\sim 1
\pm 1\ {\rm Myr}$, \citealt{h97}) provide an ideal testing ground for
these models.  \citet{h2002} found a bimodal period distribution in
this cluster for $M > 0.25 \msun$, with short- and long-period peaks,
where the long-period peak corresponds to stars which are still locked
to their circumstellar discs, preventing spin-up, whereas the stars
comprising the short-period peak have  had time to spin-up since
becoming unlocked from their discs.  The distribution appeared to be
unimodal at low-masses ($M < 0.25 \msun$) with only a short-period
peak, implying that all of the low-mass stars were fast rotators and
presumably not disc-locked.

After the accretion discs dissolve (by $20\ {\rm Myr}$ for $90\%$
  of stars, see also \citealt{hll2001}), the stars 
  are free to spin up as they contract on the PMS.  During this phase
  of the evolution, the stellar contraction dominates the angular
  momentum distribution.  Disc lifetimes can be constrained by
  examining the rotation period distributions in this age range.

As the stars approach the zero-age main sequence (at $\sim
30-100\ {\rm Myr}$ for G and K spectral types), the contraction rate
slows, and angular momentum losses begin to dominate.  These are
thought to result from magnetised stellar winds \citep{wd67} and cause
the stars to spin down gradually on the main sequence.  Previous 
studies (eg. see \citealt{st2003} for a recent  review) have shown
that most stars in young clusters ($\sim 100\ {\rm Myr}$) rotate
slowly ($v \sin i < 20\ {\rm km\ s^{-1}}$), but a minority rotate much
faster ($v \sin i \sim 200\ {\rm km\ s^{-1}}$ for the fastest
rotators).  Moving to older clusters such as the Hyades ($\sim 625\
{\rm Myr}$, \citealt{p98}), nearly all of the velocities of G and K
stars have fallen to $\la 10\ {\rm km\ s^{-1}}$.  This process must
have occurred over the age range ($\sim 200\ {\rm Myr}$) in which M34
lies.

The observations indicate that there is a large convergence in
rotation rate from the Pleiades ($\sim 125\ {\rm Myr}$,
\citealt{ssk98}) to the Hyades, but with little change
for the slowest rotators.  The change in rotation rate is also found
to be mass-dependent, in the sense that lower-mass stars require lower
angular momentum loss rates in order to reproduce the
observations \citep{bs96}.  In terms of the models, assuming solid body
rotation, in the simplest \citet{sk72} model the angular momentum
evolves as the $-1/2$ power of the age $t$ of the star.  This does not
reproduce the observed convergence in rotation rates, or the
mass-dependence.  Furthermore, the observations appear to require a
steeper early evolution than $t^{-1/2}$, followed by a flattening at
later times.

Core-envelope decoupling (eg. \citealt{s93}) has been invoked to
explain these observations.  In the core-envelope decoupling scenario,
the stars undergo differential rotation.  While the core is decoupled
from the envelope, all the angular momentum that is lost to the
outside is taken from the envelope only. This rapidly spins down the
surface layers, giving rise to the observed steep early evolution,
while the core retains a high rotation rate.  Later, as the core and
envelope are gradually recoupled, the core effectively acts as a
reservoir of angular momentum, giving rise to the observed flattening
of the rotational evolution.  These models are constrained by the
situation in the sun, which is approximately a solid body rotator,
ie. no differential rotation, so the decoupling must evolve out (see
also \citealt{ekj2002} and references therein).

Results from more comprehensive models assuming solid body rotation
are now able to provide an alternative explanation for the
observations.  For example, in the models of \citet{bfa97}, the more
rapid spin-down of faster rotators (and hence convergence in rotation
rates) can be explained in terms of their adopted angular momentum
loss law, without requiring core-envelope decoupling, and the longer
spin-down timescale for low-mass stars is explained by mass-dependent
saturation of the angular momentum losses.  In the $\alpha - \omega$
dynamo model, the mass dependency is a natural consequence of assuming
that the saturation in angular momentum losses results from saturation
of the dynamo itself.  Alternatively, \citet{giam96} have suggested
that that fully-convective stars have complex magnetic field
geometries, mostly consisting of small loops contributing little to
angular momentum losses.  The change in magnetic field geometry
toward lower-mass stars, with more closed loops, could then also give
rise to a mass-dependent saturation of the angular momentum losses.

A review of these ideas is given in \citet{b97}, based on the results
of $v \sin i$ measurements in a number of open clusters, including
$\alpha$ Persei, the Pleiades and the Hyades.

Recent work has been carried out at lower masses, eg. \citet{se2004},
who measured rotation periods for 9 very low-mass (VLM) Pleiades
members with $0.08 < M/\msun < 0.25$.  They found a lack of slow
rotators, with observed rotation periods from $\sim 0.012 - 1.7\ {\rm
  days}$, tending to increase linearly with mass.  The implications of
these results are that any angular momentum loss experienced by VLM
stars must be very small, to preserve the high rotation velocities, or
equivalently, the braking timescale of VLM stars is longer than that
of higher-mass stars.  These results lend weight to the idea that
angular momentum loss is a strong function of stellar mass, and
weakens towards the VLM regime (eg. \citealt{k97} for solar-mass stars
and \citealt{bfa97}).

Mid- to low-mass stars ($0.4 \la {\rm M}/\msun \la 1.0$) in
intermediate-age clusters such as M34 have experienced a period of
evolution after the PMS spin-up, and are thus ideal for constraining
spin-down timescales on the early main sequence.  At very low-mass, we
can also examine the late PMS angular momentum evolution.

\subsection{The survey}

We have undertaken a photometric survey in M34 using the Isaac Newton
Telescope (INT).  Our goals are three-fold: first, to establish a
catalogue of candidate low-mass M34 members, second, to study
rotation periods in a sample of these low-mass members, and third, to
look for eclipsing binary systems containing low-mass stars, to obtain
dynamical mass measurements in conjunction with radial velocities from
follow-up spectroscopy.  Such systems provide the most accurate
determinations of fundamental stellar parameters (in particular,
masses) for input to models of stellar evolution, which are poorly
constrained in this age range.  We defer discussion of our eclipsing
binary candidates to a later paper once we have obtained suitable
follow-up spectroscopy.

These observations are part of a larger photometric monitoring
survey of young open clusters over a range of ages and metalicities
(the Monitor project, \citealt{hodg06} and Aigrain et al., in prep).

The remainder of the paper is structured as follows: the observations
and data reduction are described in \S \ref{odr_section}, and the
colour magnitude diagram (CMD) of the cluster and candidate membership
selection are presented in \S \ref{memb_section}.  The method we use
for obtaining photometric periods is presented in \S
\ref{period_section}, and the resulting rotation periods and
amplitudes are discussed in \S \ref{results_section}.  Finally, we
summarise our conclusions in \S \ref{conclusions_section}.

\section{Observations and data reduction}
\label{odr_section}

Photometric monitoring data were obtained using the $2.5\ {\rm m}$
INT, with the Wide Field Camera (WFC) during a 10-night observing run
in November 2004.  This instrument provides a field of view of
approximately $34' \times 34'$ at the prime focus of the INT, over a
mosaic of four $2 {\rm k} \times 4 {\rm k}$ pixel CCDs, with $\sim
0.33''$ pixels.  Our primary target was the Orion Nebula Cluster
(ONC), which was observable for only half the night from La Palma, so
the first $\sim 4.5\ {\rm hours}$ of each night were used to observe
M34.

\begin{figure}
\centering
\includegraphics[angle=270,width=3.4in]{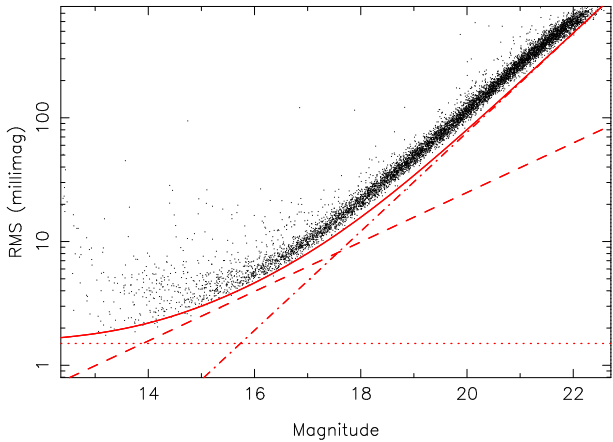}

\caption{Plot of RMS scatter per data point as a function of magnitude
for the $i$-band observations of M34, for all unblended objects with
stellar morphological classifications.  The diagonal dashed line shows
the expected RMS from Poisson noise in the object, the diagonal
dot-dashed line shows the RMS from sky noise in the photometric
aperture, and the dotted line shows an additional $1.5\ {\rm mmag}$
contribution added in quadrature to account for systematic effects.
The solid line shows the overall predicted RMS, combining these
contributions.}

\label{m34_rms}
\end{figure}

Monitoring was carried out using alternating $60\ {\rm s}$ $V$-band
and $30\ {\rm s}$ $i$-band exposures, giving an observing cadence of
$\sim 3.5\ {\rm minutes}$, over a $\sim 0.3\ {\rm sq. deg}$ region of
the cluster.  We also obtained $\sim 2 \times$ $300\ {\rm s}$
$H\alpha$ exposures per night.  Data were obtained on a total of 8 out
of the 10 nights (many of these were partial due to poor weather),
giving $275$ $V$-band and $268$ $i$-band frames after rejection of
observations influenced by technical problems (image trailing), and
$6$ $H\alpha$ frames.  Our observations are sufficient to give an RMS
per data point of $1 \%$ or better down to $i \sim 17$, with saturation
at $i \sim 13$ (see Figure \ref{m34_rms}), corresponding to a range of
spectral types for M34 members from mid-G to early-M.

For a full description of our data reduction steps, the reader is
referred to Irwin et al. (in prep).  Briefly, we used the pipeline for
the INT wide-field survey \citep{il2001} for 2-D 
instrumental signature removal (crosstalk correction, bias correction,
flatfielding, defringing) and astrometric and photometric
calibration.  We then generated the {\em master catalogue} for each
filter by stacking $20$ of the frames taken in the best conditions
(seeing, sky brightness and transparency) and running the source
detection software on the stacked image.  The resulting source
positions were used to perform aperture photometry on all
of the time-series images.  We achieved a per data point photometric
precision of $\sim 2-5\ {\rm mmag}$ for the brightest objects, with
RMS scatter $< 1 \%$ for $i \la 17$ (see Figure \ref{m34_rms}),
corresponding to $\sim 1100$ unblended stars.

Our source detection software flags as likely blends any objects
detected as having overlapping isophotes.  This information is used,
in conjunction with a morphological image classification flag also
generated by the pipeline software \citep{il2001} to allow us to
identify non-stellar or blended objects in the time-series
photometry.

Photometric calibration of our data was carried out using regular
observations of \citet{l92} equatorial standard star fields in the
usual way.  This is not strictly necessary for the purely differential
part of a campaign such as ours, but the cost of the extra telescope
time for the standards observations is negligible, for the benefits of
providing well-calibrated photometry (eg. for the production of
CMDs).

Lightcurves were extracted from the data for $\sim 14,000$ objects,
$8500$ of which had stellar morphological classifications, using our
standard aperture photometry techniques, described in Irwin et al. (in
prep).  We fit a 2-D quadratic polynomial to the residuals in each
frame (measured for each object as the difference between its
magnitude on the frame in question and the median calculated across
all frames) as a function of position, for each of the $4$ WFC CCDs
separately. Subsequent removal of this function accounted for effects
such as varying differential atmospheric extinction across each frame.
Over a single WFC CCD, the spatially-varying part of the correction
remains small, typically $\sim 0.02\ {\rm mag}$ peak-to-peak.  The
reasons for using this technique are discussed in more detail in Irwin
et al. (in prep).

For the production of deep CMDs, we stacked 120 observations in each
of $V$ and $i$, taken in good seeing and sky conditions (where
possible, in photometric conditions).  Since there were insufficient
observations taken in truly photometric conditions, the stacked frames
were corrected for the corresponding error in the object magnitudes by
comparing to a reference frame in known photometric conditions, and
using the common objects to place stacked frames on the correct
zero-point system.  The required corrections were typically $\la 0.5\
{\rm mag}$.  The limiting magnitudes, measured as the approximate
magnitude at which our catalogues are $50\%$ complete\footnote{These
  were measured by inserting simulated stars into each image, and 
  measuring the completeness as the fraction of simulated stars
  detected, as a function of magnitude.}, on these images
were $V \simeq 23.4$ and $i \simeq 22.6$.  For $H\alpha$ we stacked
all of the observations taken in sufficiently good conditions ($4$
frames), giving a limiting magnitude of $H\alpha \simeq
21.7$\footnote{The $H\alpha$ magnitudes in this paper are
  approximately equivalent to $r$-band magnitudes for a continuum
  source, and were calibrated by observing \citet{l92} standard
  stars.}  A single $600\ {\rm s}$ $r$-band frame was also taken,
with a limiting magnitude of $r \simeq 22.1$.

\section{Selection of candidate low-mass members}
\label{memb_section}

The first step in our survey is to identify the likely low-mass
cluster members.  The proper motion surveys discussed in \S
\ref{intro_section} are not suitable because they become incomplete at
the bright end of our magnitude range, eg. $V \ga 13$ assuming an
apparent distance modulus to M34 of $8.60$, estimated from Figure 8 of
\citet{jp96}.  We have therefore used colour-magnitude
diagrams (CMDs) to search for candidate low-mass members in M34.

Since we only have optical photometry available over the whole mass
range, the analysis here is preliminary and yields candidate cluster
members only.  We plan to make use of follow-up near-IR observations
and spectroscopy to improve the rejection of field stars, and will
publish a more detailed membership survey of M34 in due course.

\subsection{The $V$ versus $V - I$ CMD}
\label{cmd_section}

\begin{figure}
\centering
\includegraphics[angle=270,width=3.5in]{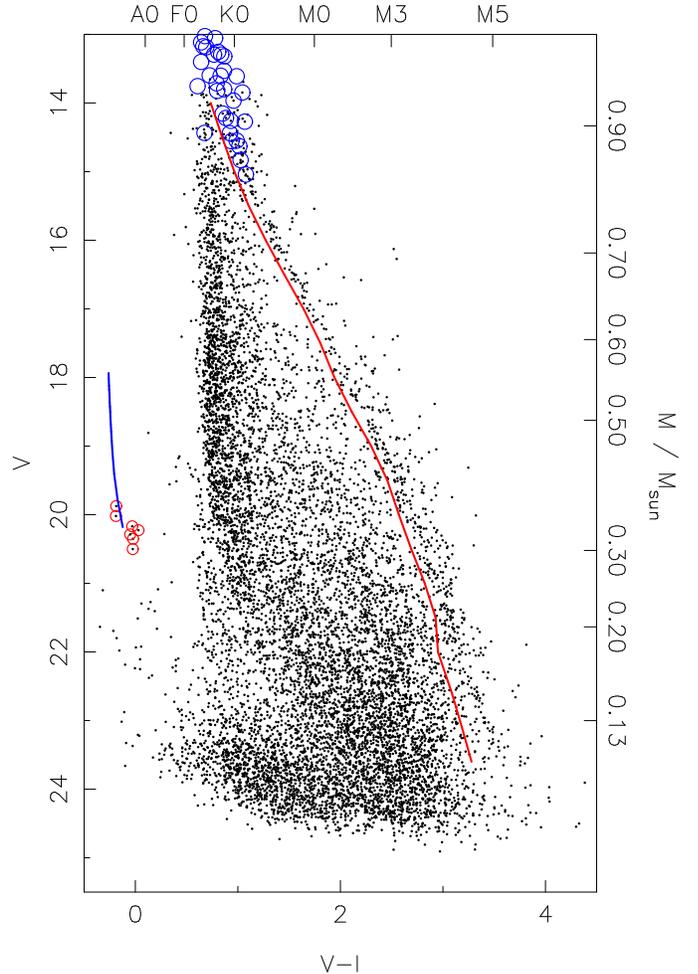}

\caption{$V$ versus $V - I$ CMD of M34 from stacked images, for all
objects with stellar morphological classification.  The cluster main
sequence is clearly visible on the right-hand side of the diagram.
The cut used to select candidate photometric members is shown as the
solid line (all objects above and to the right were selected).  The
objects in our sample assigned membership probabilities $P > 
80\%$ by \citet{jp96} are shown as open circles.  The mass scale
is from the $200\ {\rm Myr}$ NextGen model isochrone \citep{bcah98},
using our empirical isochrone to convert the $V$ magnitudes to $I$
magnitudes, and subsequently obtaining the masses from these, due to
known problems with the $V$ magnitudes from the models (see \S
\ref{cmd_section}).  The solid line and small circles at the lower
left of the diagram show the models of Bergeron et al. (private
communication) at the distance modulus of M34, for $\log g = 8.5$ to
an age of $250\ {\rm Myr}$, and our $7$ white dwarf candidates (see \S
\ref{wd_section}).}

\label{m34_cmd}
\end{figure}

A $V$ versus $V - I$ CMD of M34 was produced for selecting the
candidate cluster members, and is shown in Figure \ref{m34_cmd}.  The
INT/WFC $V$ and $i$ measurements were converted to the
Johnson-Cousins system of \citet{l92} using colour equations
derived from a large number of standard star observations, from the
INT Wide Field Survey web pages\footnote{\tt http://www.ast.cam.ac.uk/\~{}wfcsur/}:
\begin{eqnarray}
(V - I)& = &(V_{ccd} - i_{ccd})\ /\ 0.894 \\
V& = &V_{ccd} + 0.005\ (V - I) \\
I& = &i_{ccd} - 0.101\ (V - I)
\end{eqnarray}

Candidate cluster members were selected by defining an empirical main
sequence.  This was initially done `by eye' and refined from the
initial guess by computing an iterative $k \sigma$ clipped median in
$1\ {\rm mag}$ bins of $V$ magnitude, spaced at $0.5\ {\rm mag}$, down
the CMD from $V = 14$ to $24$ (the range over which the main sequence
was clearly defined on the diagram).  The result is given in Table
\ref{empms_table}.  A cut was defined by moving this line
perpendicular to the mean gradient of the main sequence, toward the
faint, blue end of the diagram, by $0.1\ {\rm mag} + k \sigma(V - I)$
with $k = 2$, where $\sigma(V - I)$ is the photometric error in $V -
I$.  The resulting curve is shown in Figure \ref{m34_cmd}.  The
choice of $k$ is somewhat arbitrary, and was made to give a good
separation between cluster members and field stars.  A greater main
sequence `width' was allowed at the faint end to account for the
increase in photometric errors.

\begin{table}
\centering
\begin{tabular}{rrrrr}
\hline
$(V-I)_0$   &$M_V$ \\
\hline
0.737 &5.4 \\
0.849 &5.9 \\
0.970 &6.4 \\
1.110 &6.9 \\
1.277 &7.4 \\
1.458 &7.9 \\
1.648 &8.4 \\
1.814 &8.9 \\
1.953 &9.4 \\
2.125 &9.9 \\
2.317 &10.4 \\
2.479 &10.9 \\
2.598 &11.4 \\
2.723 &11.9 \\
2.865 &12.4 \\
2.983 &12.9 \\
3.020 &13.4 \\
3.180 &13.9 \\
3.309 &14.4 \\
3.451 &14.9 \\
\hline
\end{tabular}

\caption{M34 empirical main sequence in dereddened $V-I$ colour and
  absolute $V$ magnitude.  These values were computed assuming a
  cluster reddening of $E(B-V) = 0.07$ and distance modulus $(M-m)_0 =
  8.38$.}
\label{empms_table}
\end{table}

Accepting all objects brighter and redder than the empirical main
sequence ensures that multiple systems (eg. binaries), which lie to
this side of the main sequence, are not excluded.  Examining the M34
CMD in Figure \ref{m34_cmd} indicates that the additional
contamination compared to applying a restrictive cut on the red side
is not significant.  Using the technique described, we found $714$
candidate members.

We also considered using the model isochrone of \citet{bcah98} for
selecting candidate members.  The model isochrone was found to be
unsuitable due to the known discrepancy between the NextGen models and
observations in the $V - I$ colour for $T_{\rm eff} \la 3700\ {\rm K}$
(corresponding here to $V - I \ga 2$).  This was examined in more
detail by \citet{bcah98}, and is due to a missing source
of opacity at these temperatures, leading to overestimation of the
$V$-band flux.  Consequently, when we have used the NextGen isochrones
to determine model masses and radii for our objects, the $I$-band
absolute magnitudes were used to perform the relevant look-up, since
these are less susceptible to the missing source of opacity, and hence
give more robust estimates.

\subsection{Completeness}
\label{comp_section}

\begin{figure}
\centering
\includegraphics[angle=270,width=3in]{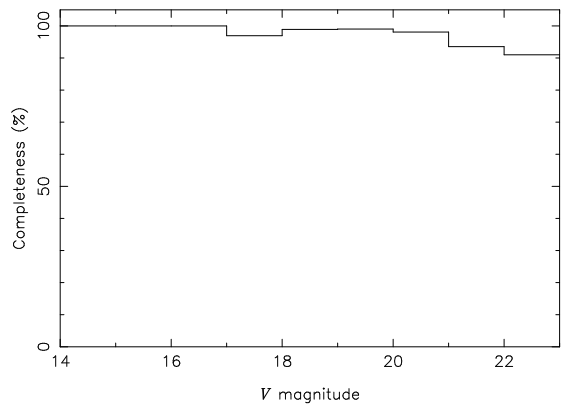}

\caption{Completeness in our source detection, measured as the
  fraction of simulated objects which were detected in each magnitude
  bin.  The diagram is plotted in the Johnson $V$ system assuming
  objects to lie on the empirically-derived cluster main sequence
  track of \S \ref{cmd_section} and Figure \ref{m34_cmd}.  The plot
  has been made only to $V = 23$, corresponding to the termination of
  the empirical main sequence here.}

\label{m34_comp}
\end{figure}

The completeness of our source detection was estimated by inserting
simulated stars as random $x,y$ positions into our images, drawing the
stellar magnitudes from a uniform distribution.  Figure \ref{m34_comp}
shows the resulting plot of completeness as a function of $V$-band
magnitude.  The completeness for objects on the M34 cluster sequence
is close to $100 \%$ up to the termination of the empirical main
sequence line at $V \sim 23$ ($I \sim 20$).

\subsection{Contamination}
\label{contam_section}

In order to estimate the level of contamination in our catalogue, we
used the Besan\c{c}on galactic models \citep{r2003} to generate a
simulated catalogue of objects passing our selection criteria at the
galactic coordinates of M34 ($l = 143.7^\circ$, $b = -15.6^\circ$),
covering the WFC FoV of $\sim 0.29\ {\rm sq.deg}$ (including gaps
between detectors).  We selected all objects over the apparent
magnitude range $13 < V < 24$, giving $7625$ stars.  The same
selection process as above for the cluster members was then applied to
find the contaminant objects.  A total of $283$ simulated objects
passed these membership selection criteria.

\begin{figure}
\centering
\includegraphics[angle=270,width=3in]{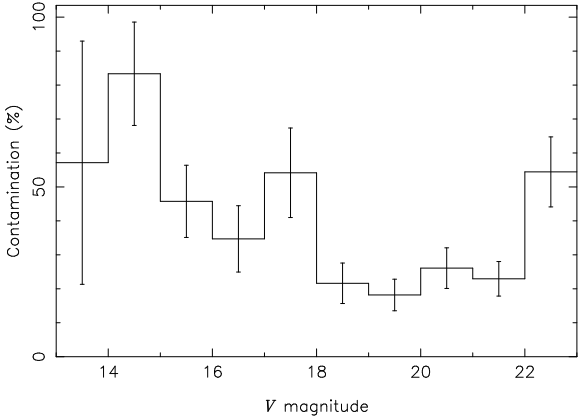}

\caption{Contamination, measured as the ratio of the calculated number
  of objects in each magnitude bin from the galactic models, to the
  number of objects detected and classified as candidate cluster
  members in that magnitude bin.  $1\sigma$ Poisson error bars are
  shown.}

\label{m34_contam}
\end{figure}

In order to account for the effects of completeness in our source
detection, the list of stellar magnitudes from the models was used to
insert simulated stars at random $x,y$ positions into our $i$-band
master image.  We then ran the source detection software on the
resulting simulated image, and kept all the inserted sources that were
detected, in order to simulate the detection process, leaving a total
of $279$ contaminant field stars.  This gave an overall contamination
level of $\sim 39 \%$, and Figure \ref{m34_contam} shows the
contamination as a function of $V$ magnitude.  We note that this
figure is somewhat uncertain due to the need to use galactic models.
Follow-up data will be required to make a more accurate estimate.

\subsection{Near-IR CMD}

\begin{figure}
\centering
\includegraphics[angle=270,width=3.0in]{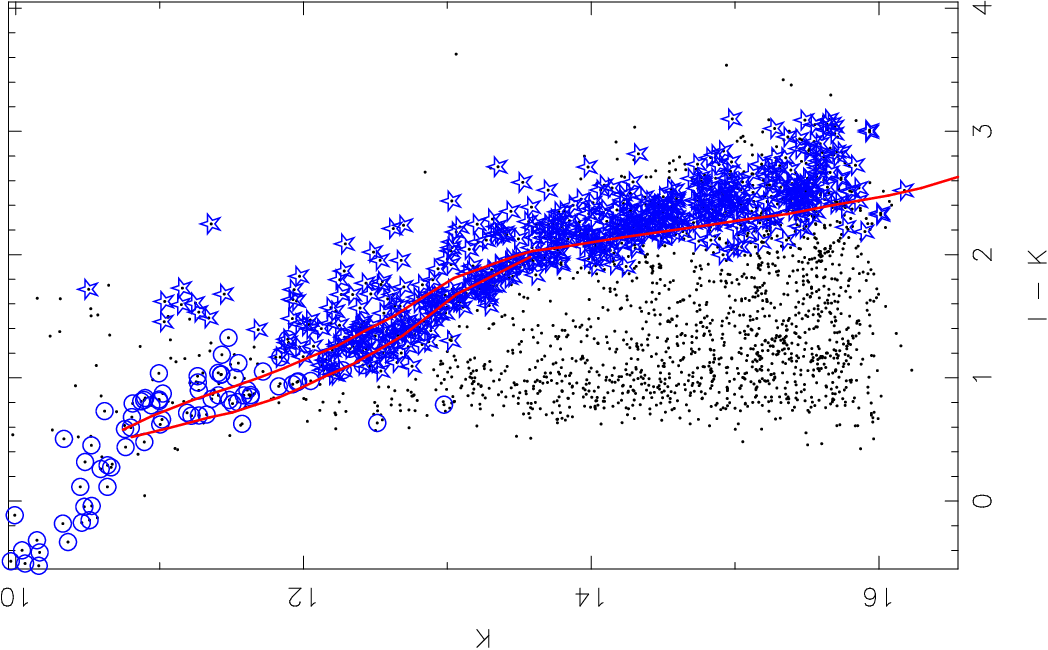}

\caption{$K$ versus $I - K$ CMD of M34, for all objects with stellar
morphological classification.  $K$ magnitudes are plotted in the CIT
system (\citealt{el82}, \citealt{el83}).  Objects classified as
members in \citet{jp96} are shown as open circles, and our
photometrically selected candidate members are shown as five-pointed
stars.  The solid lines show the $200\ {\rm Myr}$ NextGen isochrones
of \citet{bcah98} for solar metallicity, with mixing length parameters
$L_{\rm mix} = H_P$ (upper line) and $L_{\rm mix} = 1.9 H_P$ (lower
line), where the larger $L_{\rm mix}$ makes a significant difference
at high masses ($M \ga 0.6 \msun$).}

\label{m34_imk_cmd}
\end{figure}

In order to check our candidate membership assignments, we used the
$K$-band data for M34 from the Point Source Catalogue (PSC) of the
Two-Micron All-Sky Survey (2MASS) to obtain $K$-band magnitudes for
our bright sources (2MASS is complete to $K \sim 14.3$, corresponding
to $I \sim 16.5$, $V \sim 18.5$, or $\sim {\rm M2}$ spectral
types in M34).  Figure \ref{m34_imk_cmd} shows a colour-magnitude
diagram of $K$ versus $I - K$.

The locations of our photometrically-selected candidate cluster
members lie along the NextGen isochrone in Figure \ref{m34_imk_cmd},
indicating that our candidate member selection worked reasonably well
over the magnitude range covered by the 2MASS data.

\subsection{Alternative methods}

We examined alternative methods for membership selection.  The use of
colour-colour diagrams to reduce contamination was suggested, but
found to be of limited use for selection of candidate cluster members,
compared to the standard CMD method we used above.  The proper motion
of M34 is $\sim 20\ {\rm mas\ yr^{-1}}$ (eg. \citealt{jp96}).  We
examined the possibility of using photographic data-sets (in
particular the two Palomar Observatory Sky Survey epochs), but the
dispersion in the proper motions of $\sim 60\ {\rm mas\ yr^{-1}}$ for
the field objects \citep{jp96} combined with uncertainties $\sim 10\
{\rm mas\ yr^{-1}}$ at $I > 16$ are too large to allow a clean
separation of the cluster from the field for our targets. The only
suitable comparison data-set for a significant sample of new proper
motions is 2MASS, with a $\sim 5\ {\rm yr}$ baseline, ie. total motion
of $0.1''$, smaller than the RMS astrometric errors.  A proper motion
study would be feasible in $\sim 15\ {\rm years}$ from the time of
writing, when the motion would amount to $\sim 0.4''$ (which should be
measurable given the $\sim 0.1''$ RMS accuracy of 2MASS).  We can
nevertheless use proper motions for weeding out rapidly-moving
foreground objects, and intend to do this for the final membership
analysis.

In terms of angular size and density of sources, M34 would be an ideal
target for a radial velocity survey using multi-object spectroscopy on
$4$ or $8\ {\rm m}$ class telescopes.  Obtaining radial velocities
(RVs) for a large sample of our candidate members could be used to
improve the membership selection, using relatively low dispersion,
eg. $R \sim 10,000$ given the cluster RV of $\sim -8\ {\rm
  km s^{-1}}$ (estimated from Table 3 of \citealt{jp96}).

\subsection{White dwarfs}
\label{wd_section}

A number of faint, blue objects are visible in the CMD of Figure
\ref{m34_cmd}.  White dwarfs (WDs) in M34 lie in this region of the
diagram, with the WD cooling tracks of Bergeron et al. (private
communication) shown as a line on Figure \ref{m34_cmd}.  At such
young ages, we may be able to learn about the evolution of young,
massive stars from these objects.  We have selected $7$ objects in the
CMD lying in the region occupied by WDs at the distance of M34 as
candidate WD members, to be confirmed spectroscopically.

\subsection{Luminosity and mass functions}

We have calculated preliminary luminosity and mass functions using the
photometric selection.  Final versions will be published after we have
obtained sufficient additional data to more reliably determine
membership for our candidates.

\begin{figure}
\centering
\includegraphics[angle=270,width=3in]{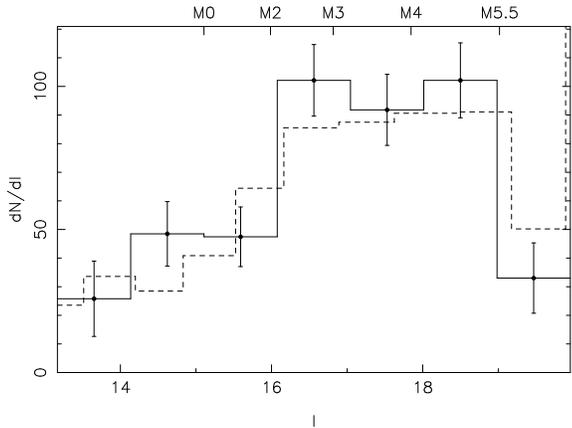}

\caption{M34 luminosity function (solid line) in apparent $I$-band
  magnitude, corrected for contamination.  Note that this function has
  {\em not been corrected for incompleteness in the source detection
  and cluster membership selection}, since the correction is very small
  over the magnitude range shown.  See also \S \ref{comp_section} and
  Figure \ref{m34_comp}.  The spectral types were determined from the
  $I$-band absolute magnitudes for young disc stars of \citet{le92}.
  The dashed line shows the solar neighbourhood luminosity function of
  \citet{rgh2002}, converted to the $I$-band using our empirical
  cluster main sequence, and normalised to give the same total number
  of objects over the $I$-band magnitude range plotted.}

\label{m34_lf}
\end{figure}

The contamination-corrected luminosity function for M34 is shown in
Figure \ref{m34_lf} and tabulated in Table \ref{m34_lf_table},
computed from our catalogue of candidate cluster members.  Note that
the plot range was chosen to correspond to the range over which our
sample is close to $100 \%$ complete (see Figure \ref{m34_comp}).  The
resulting luminosity function resembles the solar neighbourhood
luminosity function of \citet*{rgh2002}.

\begin{table}
\centering
\begin{tabular}{rrrrr}
\hline
$I_1$ &$I_2$ &$\langle I \rangle$ &${\rm dN/dI}$   &${\rm \sigma(dN/dI)}$ \\
\hline
12 &13 &12.5 &3         &4 \\
13 &14 &13.5 &18        &12 \\
14 &15 &14.5 &41        &11 \\
15 &16 &15.5 &49        &10 \\
16 &17 &16.5 &101       &12 \\
17 &18 &17.5 &92        &12 \\
18 &19 &18.5 &101       &13 \\
19 &20 &19.5 &33        &12 \\
20 &21 &20.5 &-3        &3 \\
\hline
\end{tabular}

\caption{Tabulated contamination-corrected M34 luminosity function (as
  shown in Figure \ref{m34_lf}) in $1\ {\rm mag}$ bins.
  $I_1$ and $I_2$ denote the start and end of the bin in $I$, $\langle
  I \rangle$ the central $I$ value of the bin, and ${\rm
  \sigma(dN/dI)}$ gives the estimated uncertainty from Poisson
  counting errors.}
\label{m34_lf_table}
\end{table}

\begin{figure}
\centering
\includegraphics[angle=270,width=3in]{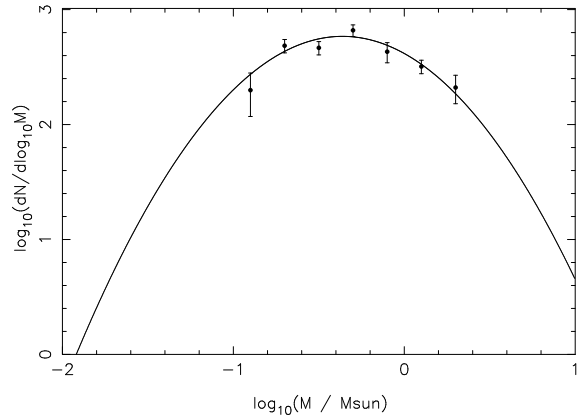}

\caption{M34 mass function in ${\rm dN/dlogM}$, corrected for
  contamination, made using the models of \citet{bcah98} to calculate a
  mass for each object from the $I$ magnitude.  The solid line shows a 
  log-normal fit to ${\rm dN/dM}$, ie. a quadratic fit to $\log(${\rm
  dN/dlogM}$)$, with parameters $M_0 = 0.44\ \msun$ (the
  mass at maximum ${\rm dN/dM}$), and $\sigma = 0.66$.
  We have used 
  the catalogue of M34 proper motion members from \citet{jp96} to
  populate the two highest-mass bins.  Note that the projected areas
  covered by these surveys are similar, but the centres are offset by
  a few arcminutes, so the actual coverage of M34 members may differ,
  which will affect the results for the upper two bins.}

\label{m34_mf}
\end{figure}

The mass function for M34 has been computed over the range $0.1 \la
M/\msun \la 2.5$ covered by the available data (our sample plus proper
motion members of M34 from \citealt{jp96}) with close to $100 \%$
completeness.  The lower limit results from the drop in completeness
at $I \simeq 20$ for our sample, and the upper limit from the upper
mass limit of the models of \citet{bcah98}.  The result is shown in
Figure \ref{m34_mf}.  In order to produce these distributions, the
observed number counts from our survey were corrected for field star
contamination using the Besan\c{c}on model star counts of \S
\ref{contam_section}.  For the data from \citet{jp96}, we chose only
those objects with membership probabilities $P > 80\ \%$.  The masses
were computed from the $I$-band magnitudes using the models of
\citet{bcah98}, using a linear extrapolation from the high-mass end
for the brightest objects from \citet{jp96}.

We find that a log-normal mass distribution of the form:
\begin{equation}
{\rm dN/dlogM} \propto \exp\left[-{\left(\log M - \log
    M_0\right)\over{2 \sigma^2}}\right]
\end{equation}
is a good fit to the data in ${\rm dN/dlogM}$, parameterized by a
mass at maximum ${\rm dN/dlogM}$ of $M_0 = 0.44 \pm 0.31\ \msun$
and $\sigma = 0.66 \pm 0.08$.  These are in relatively good agreement
with the values of $M_0 \sim 0.3\ \msun$ and $\sigma \sim 0.5$ derived
by \citet{m2005} for a sample of three young open clusters.  It
should be noted that our survey covered only the central part of the
cluster\footnote{At a distance of $470\ {\rm pc}$, $1\ {\rm pc} \sim
  0.12^\circ$, and our survey covers a radius of $\sim 0.2^\circ$,
  similar to typical cluster core radii of a few ${\rm pc}$.},
and it is likely that the low-mass members are distributed 
with a larger core radius than the higher-mass members, implying a
bias toward higher masses.  Unfortunately our present survey does not
have sufficient sky coverage to resolve this problem.  It should also
be noted that these conclusions were made on the basis of measurements
over a very small range in mass.  More data at higher or lower masses
would improve the situation and allow us to examine the mass function
in greater detail.

\begin{table}
\centering
\begin{tabular}{rrrrr}
\hline
$\log_{10} M_1$ &$\log_{10} M_2$ &$\langle \log_{10} M \rangle$ &${\rm dN/dlogM}$   &${\rm \sigma(dN/dlogM)}$ \\
\hline
-1.0 &-0.8 &-0.9 &642 &263 \\
-0.8 &-0.6 &-0.7 &1046 &139 \\
-0.6 &-0.4 &-0.5 &633 &84 \\
-0.4 &-0.2 &-0.3 &567 &65 \\
-0.2 & 0.0 &-0.1 &233 &46 \\
 0.0 & 0.2 & 0.1 &109 &15 \\
 0.2 & 0.4 & 0.3 & 54 &15 \\
\hline
\end{tabular}

\caption{Tabulated M34 mass function in ${\rm dN/dlogM}$.  $M_1$ and
  $M_2$ denote the start and end of the bin in $M$, $\langle 
  \log_{10} M \rangle$ the central $\log_{10} M$ value of the bin, and
  ${\rm \sigma(dN/dlogM)}$ gives the estimated uncertainty from
  Poisson counting errors.}
\label{m34_mf_table}
\end{table}

\section{Period detection}
\label{period_section}

\subsection{Method}
\label{method_section}

The limited time baseline of only $10\ {\rm nights}$ for our
observations is insufficient to make a detailed study of stellar
rotation periods $\ga 10\ {\rm days}$.  However, our observations
have relatively good sensitivity to short-period ($\la 10\ {\rm days}$
for example) systems so a careful examination of the data for
rotation and other periodic variability has been carried out.

Variable objects were selected using least-squares fitting of sine
curves to the time series $m(t)$ (in magnitudes) for all sources in
the CCD images, using the form:
\begin{equation}
m(t) = m_{dc} + \alpha \sin(\omega t + \phi)
\label{sine_eqn}
\end{equation}
where $m_{dc}$ (the DC lightcurve level), $\alpha$ (the amplitude) and
$\phi$ (the phase) were free parameters at each value of $\omega$ over
an equally-spaced grid of frequencies, corresponding to periods from
$0.005 - 20\ {\rm days}$.  The lower period limit was chosen to
correspond to the cadence of our observations, and the upper limit to
the observing window, estimating that we still have a chance of
recovering the correct period after observing only half a cycle (see
also \S \ref{sim_section}).  The output of this procedure is a
`least-squares periodogram', with the best-fitting period being the
one giving the smallest reduced $\chi^2$.

All of the lightcurves were processed in this manner.  In order to
distinguish the variable objects from the non-variable objects, we
used the reduced $\chi^2$, but evaluated by subtracting a smoothed,
phase-folded version of each lightcurve at the best-fitting period.
The smoothing was carried out using median filtering over a $51$
data-point window, followed by a linear boxcar filter over an $11$
data-point window to smooth out any high-frequency features.  This
procedure accounts for any non-sinusoidal (but otherwise periodic)
features in the lightcurve, and we have found empirically that it is
capable of detecting somewhat non-sinusoidal variables such as
detached eclipsing binaries and even the transiting extrasolar planet
OGLE-TR-56b \citep{k2003} from the published lightcurve data.

Periodic variable lightcurves were selected by evaluating the change
in reduced $\chi^2$:
\begin{equation}
\Delta \chi^2_\nu = \chi^2_\nu - \chi^2_{\nu,smooth} > 0.4
\end{equation}
where $\chi^2_\nu$ is the reduced $\chi^2$ of the original lightcurve
with respect to a constant model, and $\chi^2_{\nu,smooth}$ is the
reduced $\chi^2$ of the lightcurve with the smoothed, phase-folded
version subtracted.  We used a simple cut in $\Delta \chi^2_\nu$
rather than $\Delta \chi^2_\nu / \sigma(\chi^2)$ (where
$\sigma(\chi^2)$ is the RMS of the $\chi^2$ measure, depending only on
the number of data points) because our error estimates are somewhat
unreliable, and therefore $\sigma(\chi^2)$ underestimates the true
dispersion of the $\chi^2_\nu$ value for a featureless lightcurve.

\begin{figure}
\centering
\includegraphics[angle=270,width=3in]{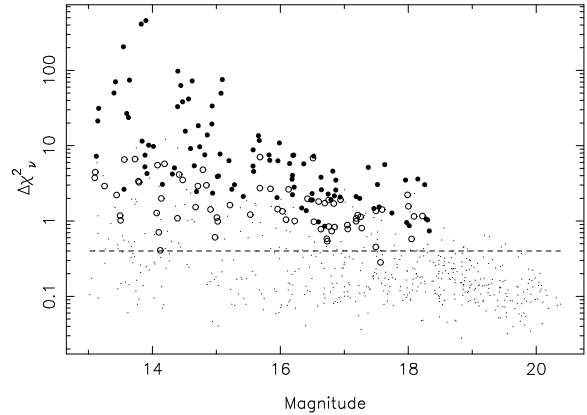}

\caption{$\Delta \chi^2_\nu$ as a function of $i$-band magnitude for
  candidate M34 cluster members.  The horizontal line shows the threshold
  at $\Delta \chi^2_\nu > 0.4$, with solid circles indicating objects
  with clear periodic variability, open circles indicating the
  ambiguous cases, and small points indicating objects without periodic
  variability (non-variable lightcurves or those showing irregular
  variability).}

\label{delchisq_figure}
\end{figure}

The threshold of $\Delta \chi^2_\nu > 0.4$ was chosen as follows.  All
of the $714$ lightcurves of candidate cluster members were examined by
eye, looking for variability, and placing the objects into four bins:
objects with clear periodic variability, objects with clear
non-periodic variability, ambiguous cases, and non-variable
lightcurves.  The numbers of objects falling in each bin were $102$,
$8$, $67$ and $537$, respectively.  Figure \ref{delchisq_figure} shows
these objects on a diagram of $\Delta \chi^2_\nu$ as a function of
magnitude.

The threshold was chosen to select at least all of the clear periodic
variables and the more significant of the ambiguous lightcurves.
Applying the threshold gave $283$ detections, after also removing
objects lying on bad pixels for more than $10 \%$ of their lightcurve
points, non-stellar objects (by morphological classification) and any
object with data points spanning a range of less than $1/2$ of the
orbital phase.  Note that objects flagged as blended were not removed
since a large fraction of these were sufficiently uncontaminated to
still be used.  Only three of the lightcurves classified as ambiguous
were rejected, and these were visually confirmed to be poor detections
that were not obviously variable.

After applying the threshold, the selected lightcurves were examined
independently by eye, to select the final sample included here.  A
total of $105$ lightcurves were selected, with the remainder appearing
non-variable or too ambiguous to be included.

Our periods derived from the selection procedure described were
refined by using both the $V$ and $i$ band lightcurves simultaneously, 
fitting a separate set of coefficients $m_{dc}$, $\alpha$ and $\phi$
in each band, but combining the $\chi^2$ values to produce a
periodogram taking both bands into account.  We have not used this
method for the detection process itself because it is difficult to
simulate for evaluating completeness, necessitating assuming a
relation between the $V$ and $i$-band amplitudes and magnitudes for
the simulated objects.  Furthermore, at the faint end, where the
detection begins to become incomplete, the signal to noise in the
$V$-band is very poor for cluster members, so using the $V$-band does
not give any improvement.

\subsection{Simulations}
\label{sim_section}

The large amount of human involvement in our selection procedure
is difficult to simulate in an unbiased manner.  Nevertheless, we have
attempted to evaluate our selection biases by performing Monte Carlo
simulations, injecting sinusoidal modulations into non-variable M34
lightcurves, selected by requiring $\chi^2_\nu <
\langle\chi^2_\nu\rangle + 3 \sigma(\chi^2_\nu)$ (calculated using
robust MAD -- median of absolute deviations from the median --
estimators in $0.5\ {\rm mag}$ bins) to remove the most variable
objects.  These were then subjected to exactly the same selection
procedure as the real lightcurves, detailed in \S
\ref{method_section}, including examination by eye.

In order to reduce biases in the detection process resulting from
`knowing' that the modulations are real, modulations were not
inserted into a fraction ($1/6$) of the lightcurves shown to the
human.  A larger fraction ($1/2$) would be more realistic but
would increase the already rather large number of lightcurves that
must be examined.

Each simulation was run following uniform distributions in $\log_{10}\
{\rm period}$ from $0.1$ to $20\ {\rm days}$, and a uniform
distribution in mass from $1.0$ to $0.1\ {\msun}$, the second
satisfied by choosing lightcurves in the correct magnitude range for
each $0.1\ \msun$ bin.  Phases $\phi$ were chosen at random in $0 \le
\phi < 2\pi$, and the exercise was repeated for three typical
amplitudes, $0.01\ {\rm mag}$, $0.02\ {\rm mag}$ and $0.05\ {\rm
  mag}$, corresponding approximately to the range of values for M34
rotators that we detected.  $\sim 2000$ periodic variable objects were
simulated in each amplitude bin.

\begin{figure}
\centering
\includegraphics[angle=270,width=3in]{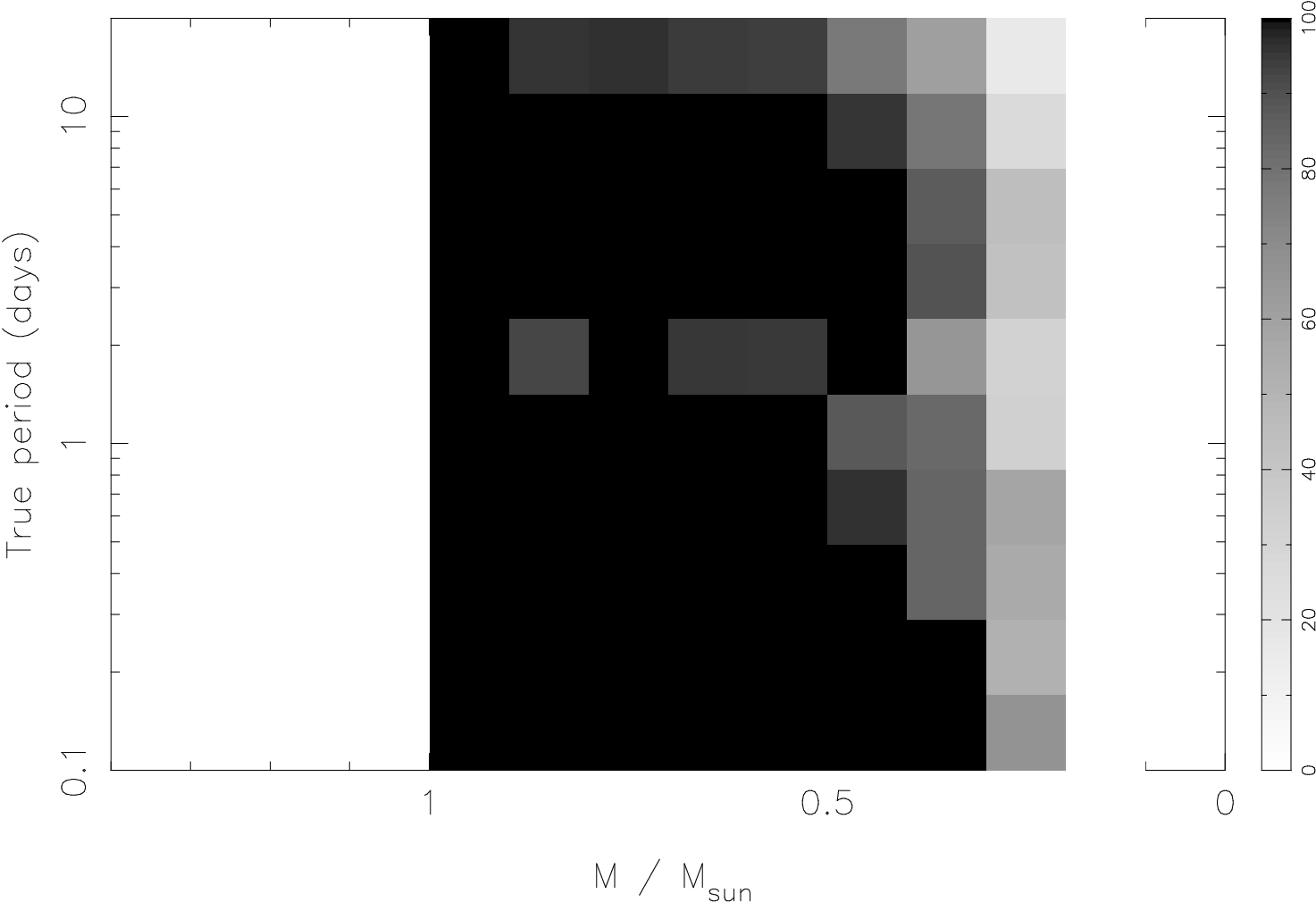}
\includegraphics[angle=270,width=3in]{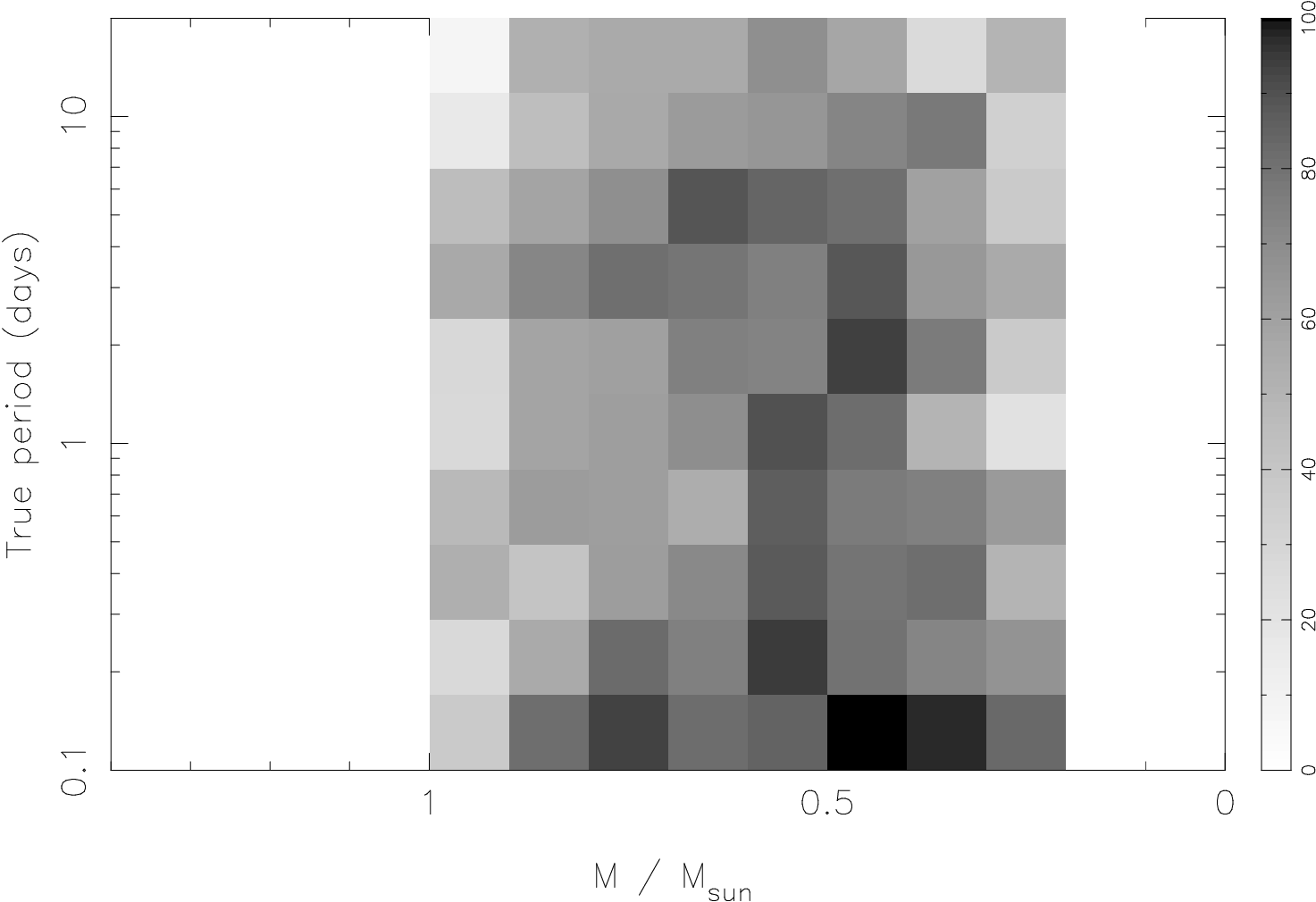}
\includegraphics[angle=270,width=3in]{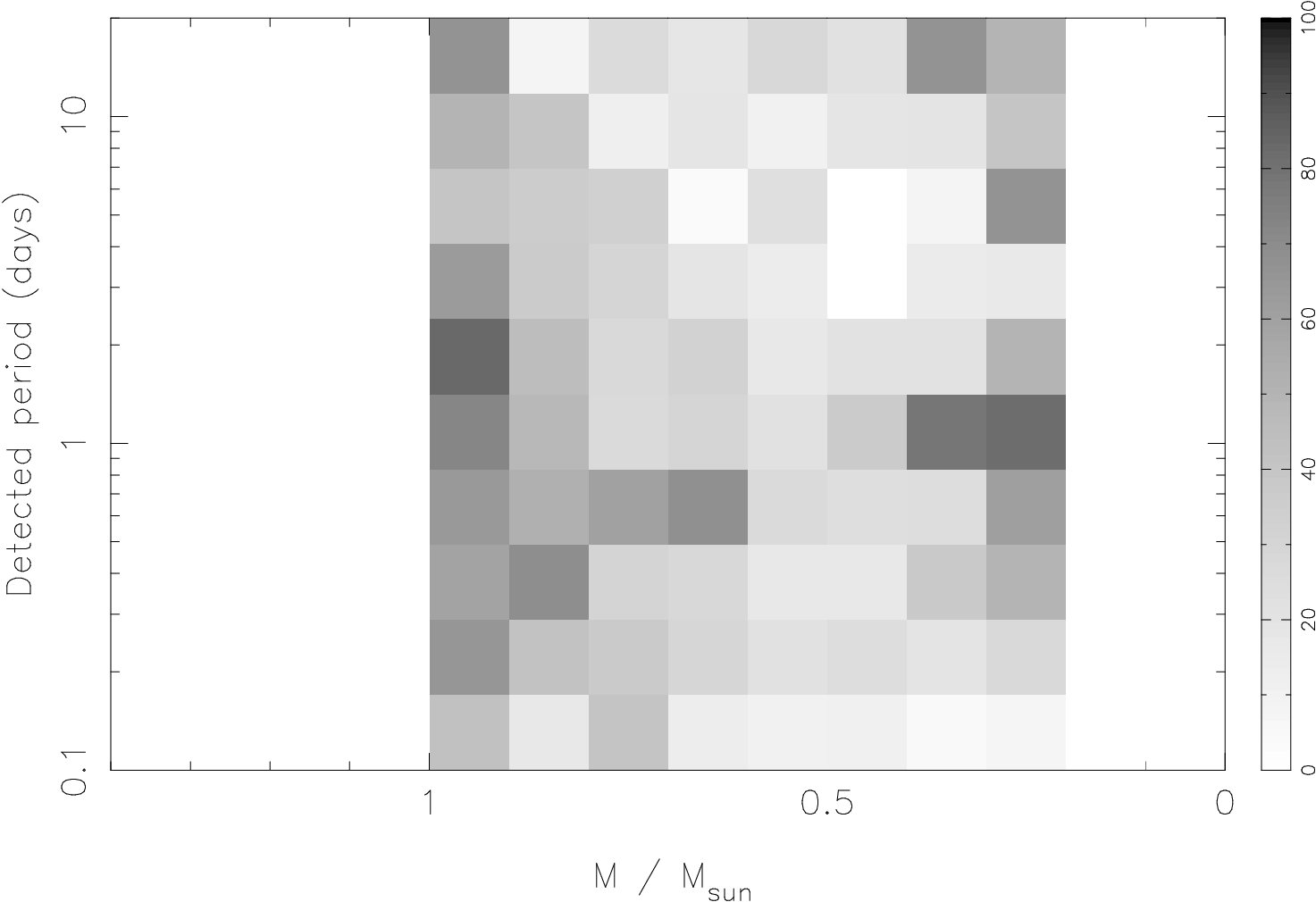}

\caption{Results of the simulations for $0.02\ {\rm mag}$ amplitude as
  greyscale maps, where black corresponds to $100 \%$ and white to $0
  \%$.  The simulated region covered $0.2 < {\rm M}/\msun < 1.0$ in
  order to be consistent with the M34 sample.  {\bf Top panel}:
  completeness as a function of real (input) period.  {\bf Centre
  panel}: Reliability of period determination, plotted as the fraction
  of objects with a given true period, detected with the correct
  period (defined as differing by $< 20\%$ from the true period).
  {\bf Bottom panel}: Contamination, plotted as the fraction of
  objects with a given detected period, having a true period differing
  by $> 20\%$ from the detected value.}

\label{m34_sim_results}
\end{figure}

The results of the simulations are shown in Figure
\ref{m34_sim_results} as greyscale diagrams of completeness,
reliability and contamination as a function of period and stellar
mass.  Broadly, our period detections are close to $100 \%$ complete
for $P < 10\ {\rm days}$ and $M \ga 0.4\ \msun$.

\begin{figure}
\centering
\includegraphics[angle=270,width=3in,bb=59 107 581 630,clip]{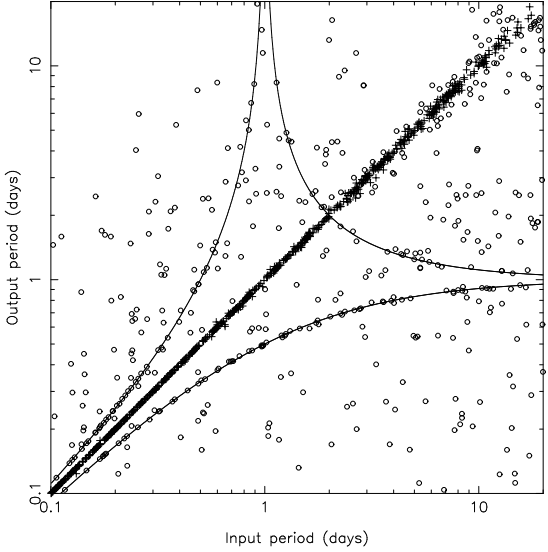}

\caption{Detected period as a function of actual (input) period for our
  simulations.  Objects plotted with crosses had fractional period
  error $< 10\%$, open circles $> 10\%$.  The straight line represents
  equal input and output periods.  The curved lines are the loci of
  the $\pm 1\ {\rm day^{-1}}$ aliases from equation (\ref{beat_eqn}).
  The majority of the points fall either on (or close to) the line of
  equal periods, or on one of the curves consistent with $1\ {\rm
  day^{-1}}$ aliasing.}

\label{m34_periodcomp}
\end{figure}

As with all ground-based observations, our period determinations
suffer problems of aliasing.  The most serious of these is at a
frequency spacing of $1\ {\rm day^{-1}}$ corresponding to the
observing gaps during the day, and leads to periodogram peaks at the
`beat' periods
\begin{equation}
1 / P_{beat} = 1 / P \pm 1
\label{beat_eqn}
\end{equation}
where $P$ is the true period.  Figure \ref{m34_periodcomp} illustrates
this effect using our simulated lightcurves.  Using both $V$ and $i$
bands to refine the period estimates mitigates the effect slightly
(since the observations were not precisely simultaneous), but this has
not been accounted for in the simulations.

\subsection{Detection rate and reliability}

\begin{figure}
\centering
\includegraphics[angle=270,width=3.5in]{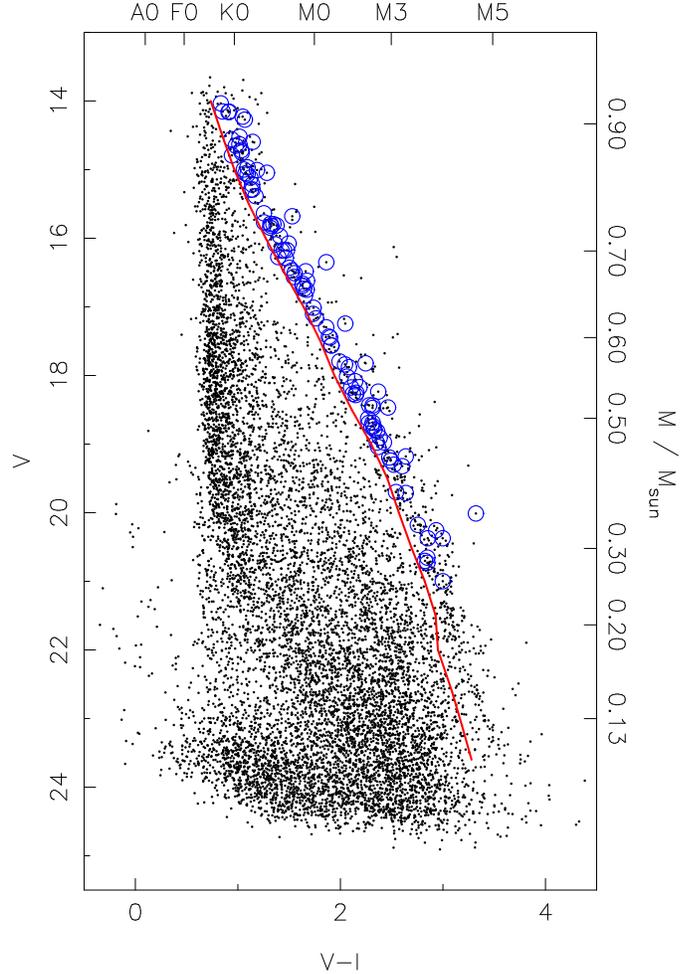}

\caption{$V$ versus $V - I$ CMD of M34, for all objects with stellar
morphological classification, as Figure \ref{m34_cmd}, showing all
$105$ candidate cluster members with detected periods (open circles).}

\label{m34_cands_on_cmd}
\end{figure}

The locations of our detected periodic variable candidate cluster
members on a $V$ versus $V-I$ CMD of M34 are shown in Figure
\ref{m34_cands_on_cmd}.  The diagram indicates that the majority of
the detections lie on the single-star cluster main sequence, as would
be expected for rotation in cluster stars as opposed to, say,
eclipsing binaries.

\begin{figure}
\centering
\includegraphics[angle=270,width=3in]{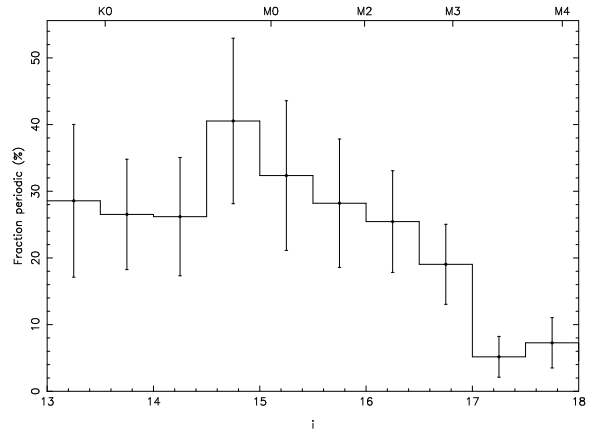}

\caption{Fraction of candidate cluster members detected as periodic
  variables, plotted as a function of magnitude.  This distribution
  has not been corrected for incompleteness in the period detections,
  so the decaying portion of the histogram from $i \sim 15$ is likely
  to be an incompleteness effect resulting from the gradual increase
  in the minimum amplitude of variations we can detect, as a function
  of increasing magnitude (and hence increasing noise, see Figure
  \ref{m34_rms}), and not a real decline.}

\label{m34_fracper}
\end{figure}

Figure \ref{m34_fracper} shows the fraction of cluster members with
detected periods as a function of $i$ magnitude.  The rise from the
bright end towards lower masses indicates that $M$-dwarfs may show
more spot-related rotational variability within our detection limits
than $K$ or $G$-dwarfs.  This would be consistent with an increase in
spot coverage moving to later spectral types.  The decaying portion of
the histogram from $i \sim 15$ is likely to be an incompleteness
effect resulting from the gradual increase in the minimum amplitude of
variations we can detect (corresponding to the reduction in
sensitivity moving to fainter stars, see Figure \ref{m34_rms}) and not
a real decline.

The phase-folded lightcurves of all our rotation candidates are shown
in Figure \ref{m34_cand_lc}, and their properties summarised in Table
\ref{m34_cand_table} (see Appendix \ref{lc_section}).

\subsection{Non-periodic objects}

The population of objects rejected by the period detection procedure
described in \S \ref{period_section} was examined, finding that the
most variable population of these lightcurves (which might correspond
to non-periodic or semi-periodic variability) was contaminated by a
number of lightcurves exhibiting various uncorrected systematic
effects.  It is therefore difficult to quantify the level of
non-periodic or semi-periodic variability in M34 from our data.
Qualitatively however, there appear to be very few of these
variables, and examining the lightcurves indicated only $\sim 3$
obvious cases, which strongly resembled eclipses (either planetary
transits or eclipsing binaries), and will be the subject of a later
Monitor project paper.

\section{Results}
\label{results_section}

\subsection{Periods and rotational velocities}
\label{perioddist_section}

\begin{figure*}
\centering
\includegraphics[angle=270,width=3in]{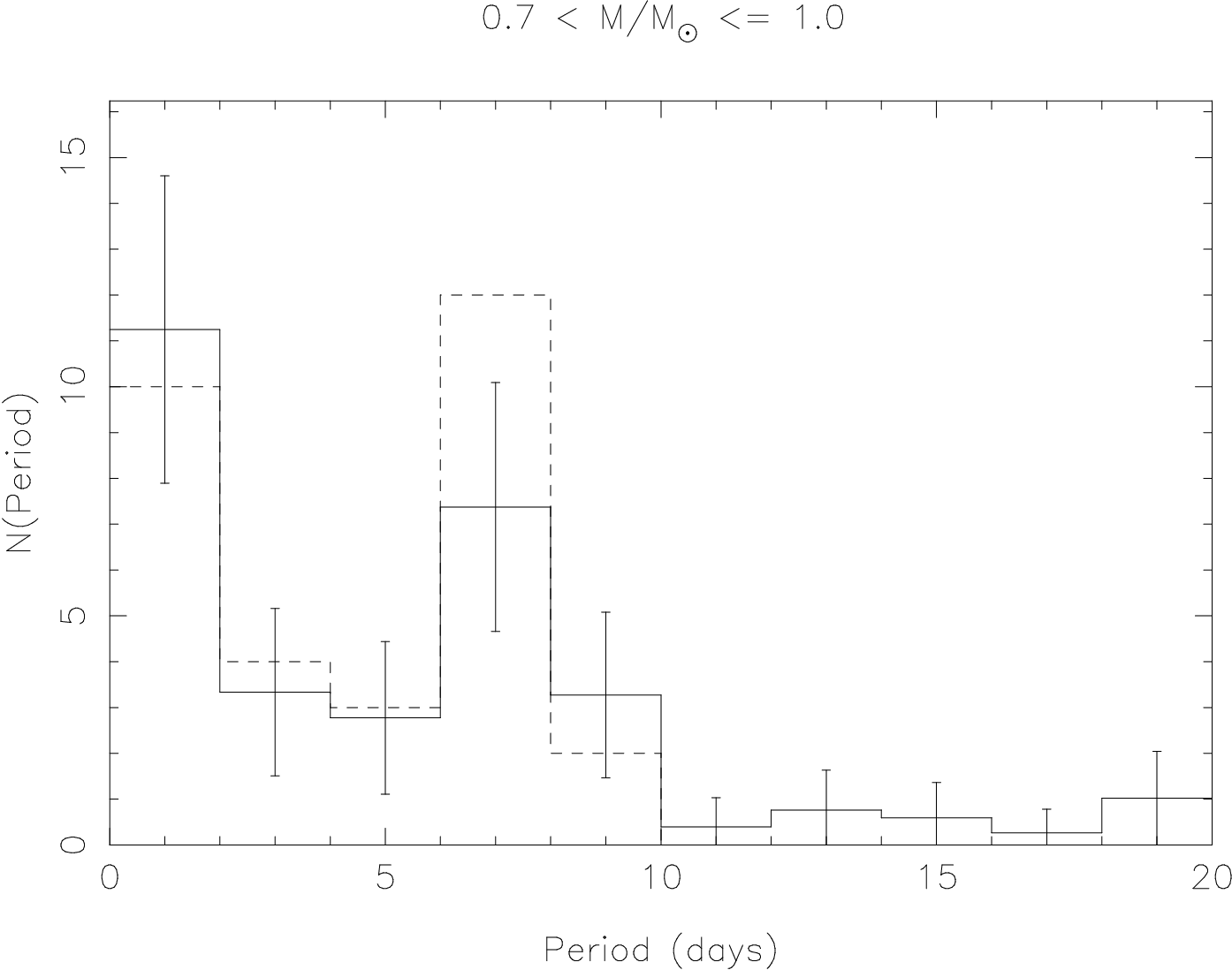}
\includegraphics[angle=270,width=3in]{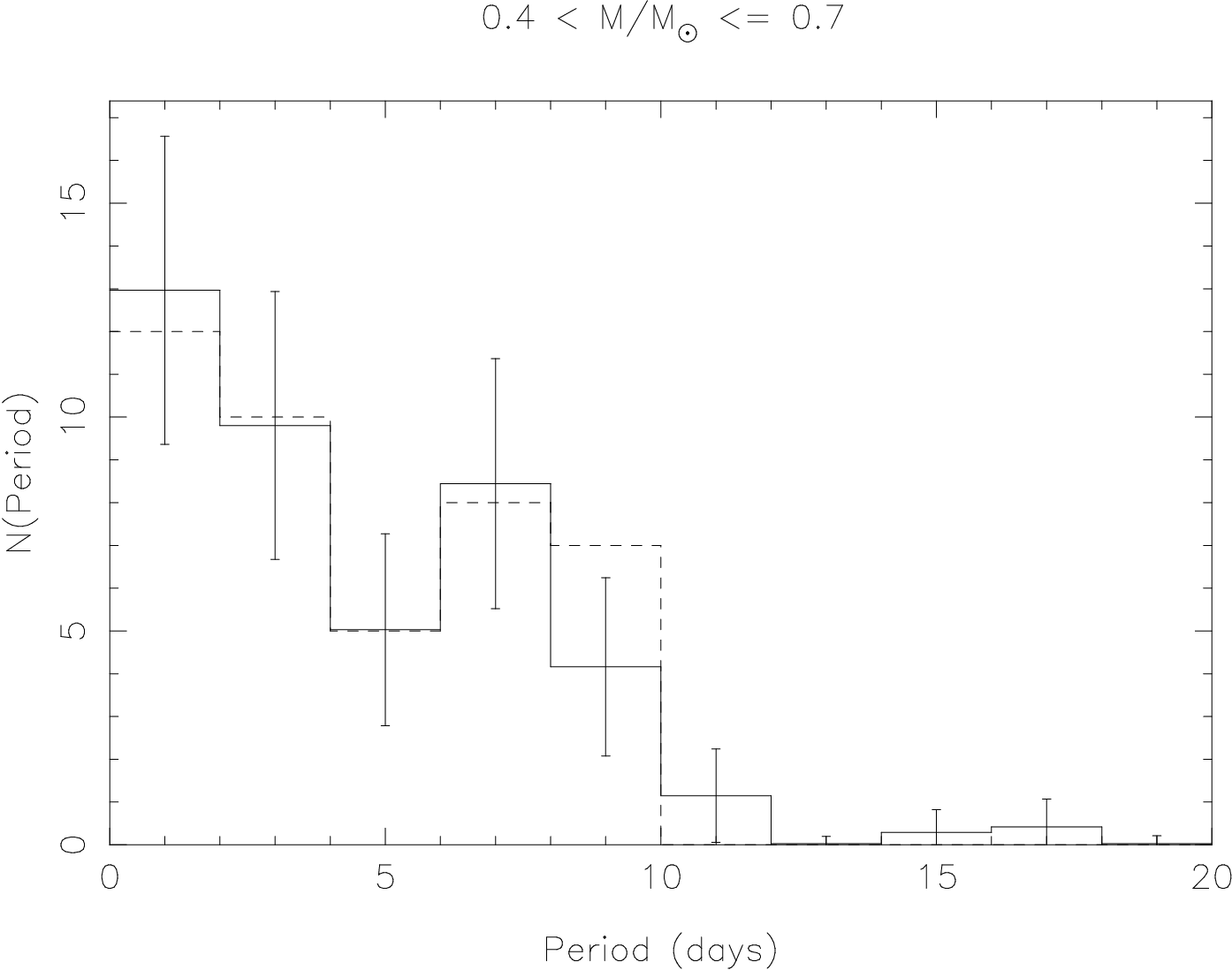}
\includegraphics[angle=270,width=3in]{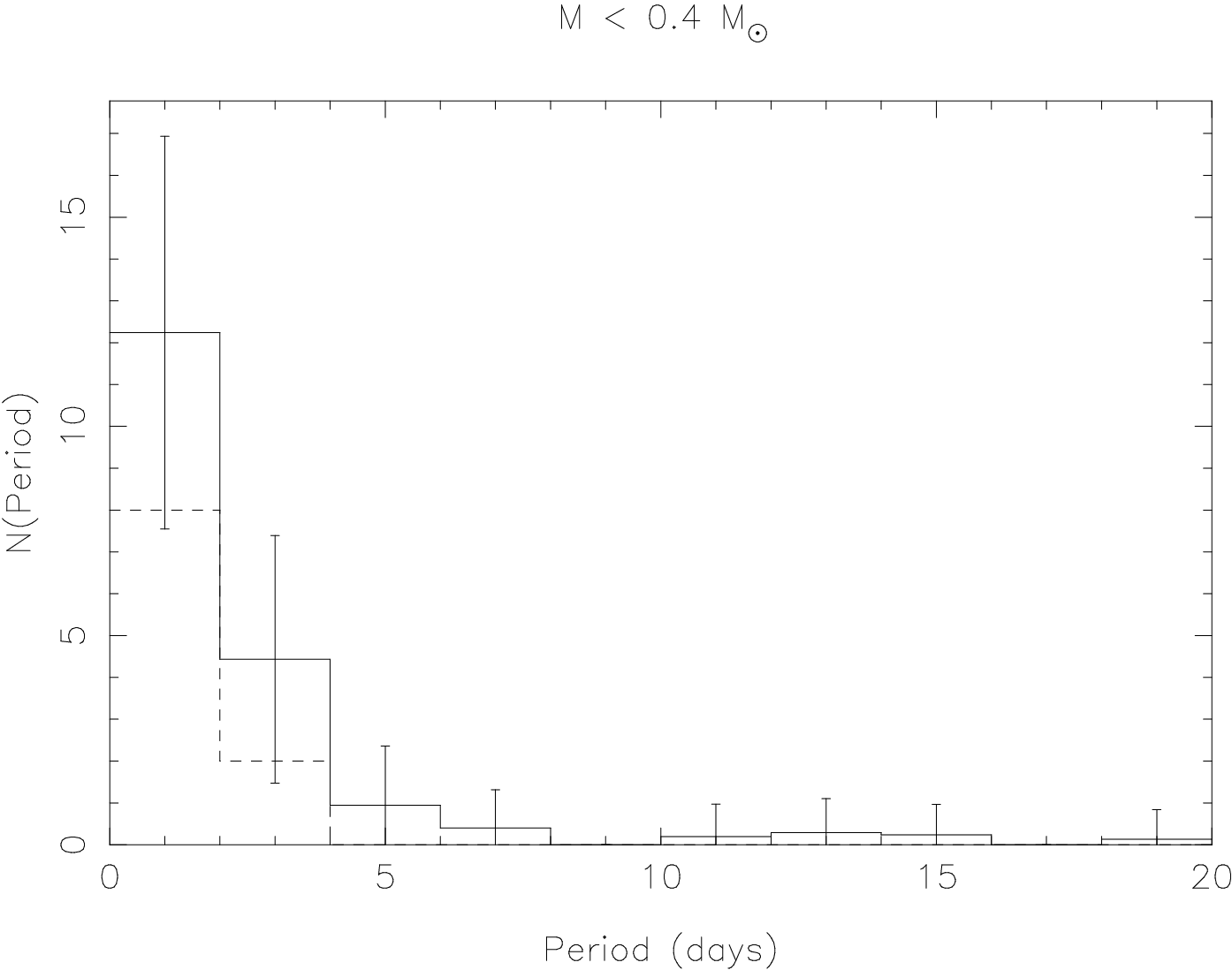}
\includegraphics[angle=270,width=3in]{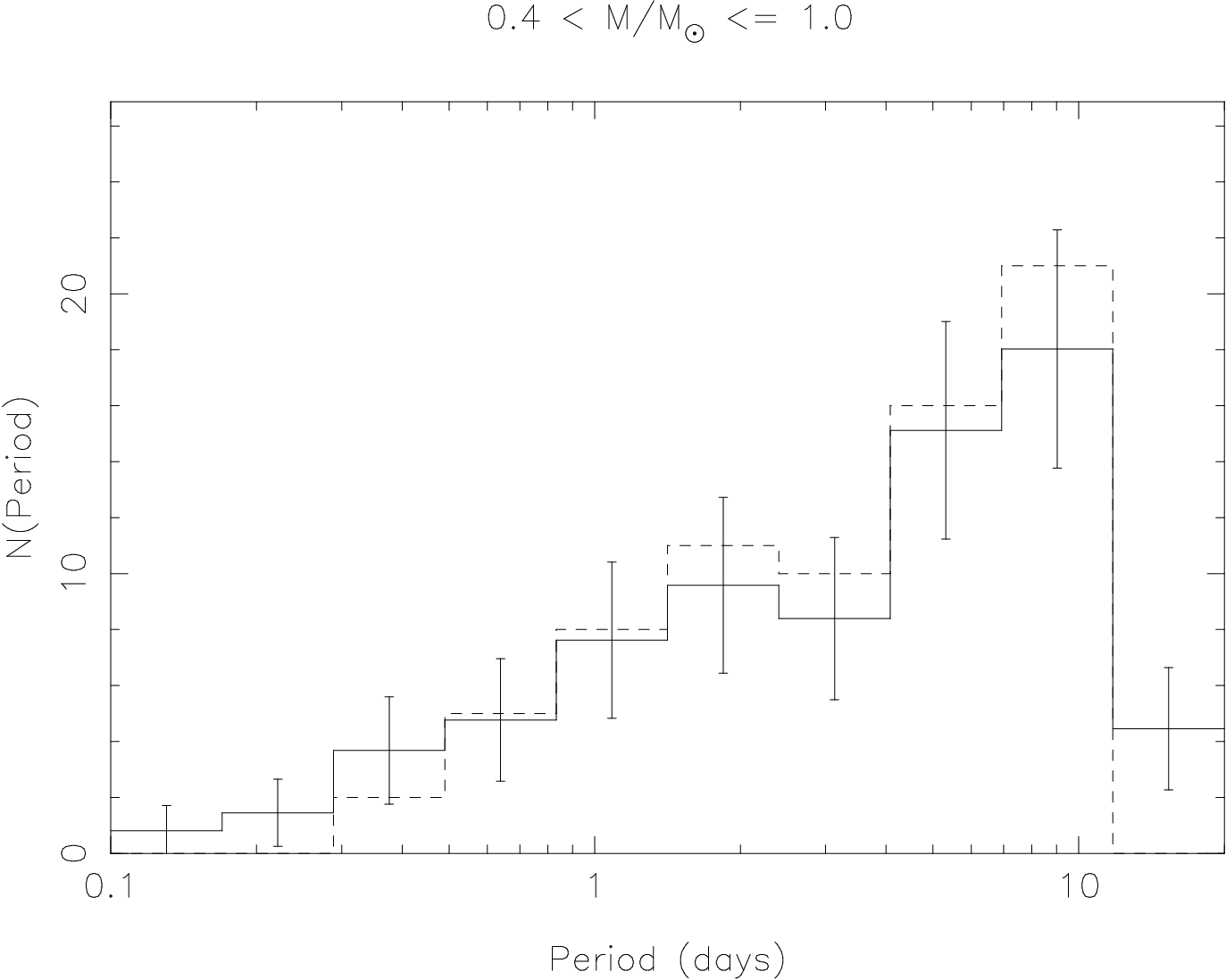}

\caption{Period distributions for objects classified as possible
  photometric members, in three mass bins: $0.7 \le M/\msun < 1.0$
  (corresponding roughly to late-G and K spectral types),
  $0.4 \le M/\msun < 0.7$ (early-M) and $M < 0.4\ \msun$ (late-M and
  VLM).  The bottom right panel shows the period distribution for $0.4
  \le M/\msun < 1.0$ plotted in $log_{10}$ period, indicating that the
  apparent bimodality in this mass range is probably not significant.
  The dashed lines show the measured period distributions, and
  the solid lines show the results of attempting to correct for
  incompleteness and reliability, as described in the text.  Objects
  with periods flagged as unreliable in Table \ref{m34_cand_table}
  have been excluded.}

\label{m34_perioddist}
\end{figure*}

\begin{figure}
\centering
\includegraphics[angle=270,width=3in]{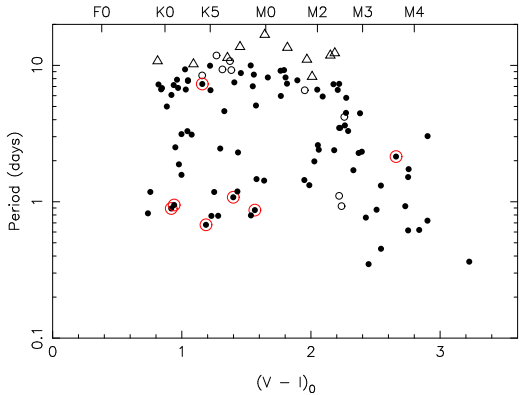}

\caption{Plot of rotation period as a function of dereddened $V-I$
  colour for M34.  X-ray sources from \citet{s2000} are overlaid with
  open circles.  Open symbols indicate objects flagged `a' (open
  circles) and `b' (open triangles) in Table \ref{m34_cand_table}.}
\label{m34_pcd}
\end{figure}

\begin{figure}
\centering
\includegraphics[angle=270,width=3in]{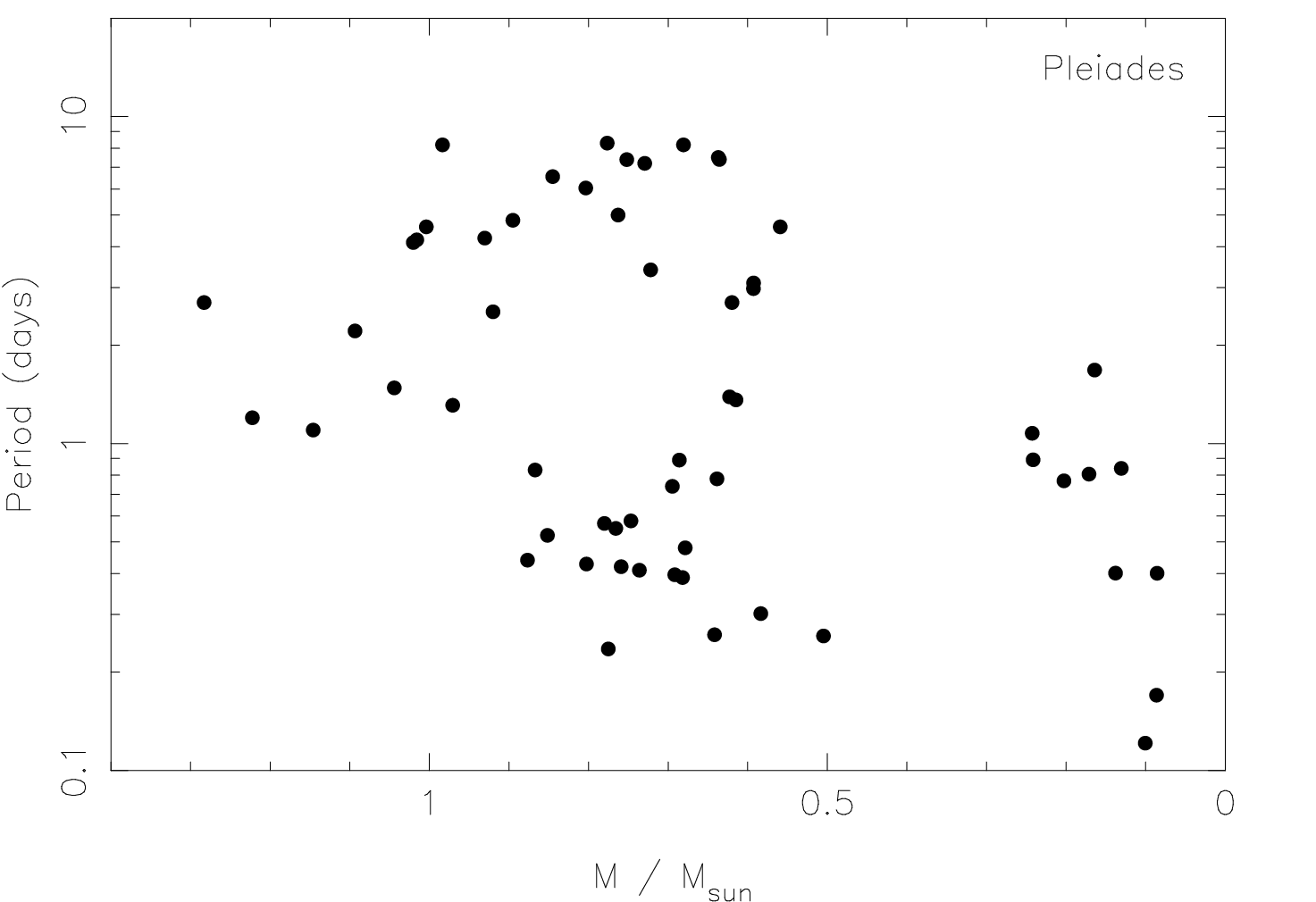}
\includegraphics[angle=270,width=3in]{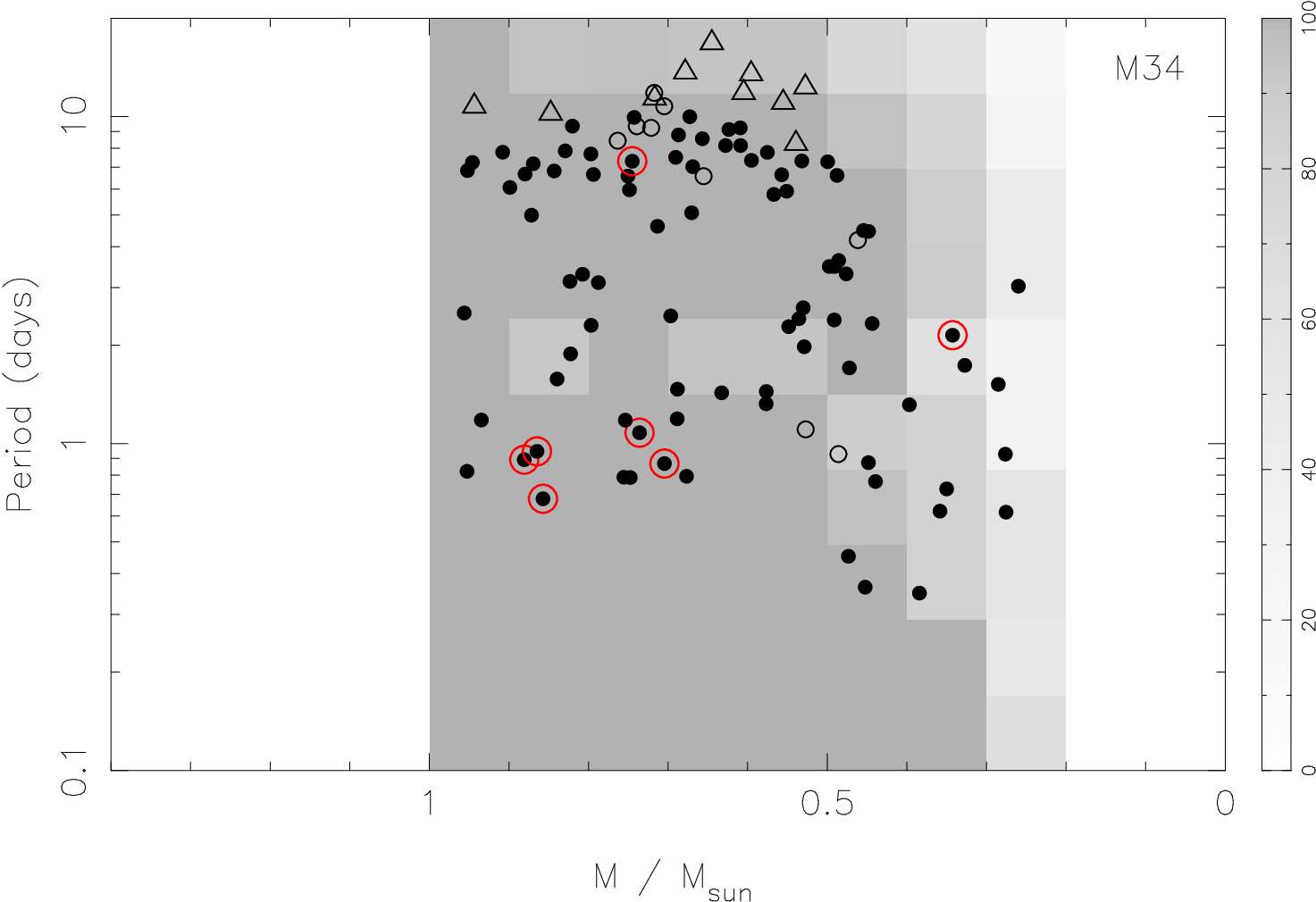}

\caption{Plot of rotation period as a function of mass for the
  Pleiades (top) and M34 (bottom).  Ages: $125\ {\rm Myr}$
  \citep{ssk98} and $200\ {\rm Myr}$.  Masses were interpolated from
  the model isochrones of \citet{bcah98} using the $I$-band magnitudes
  (two Pleiades objects without this information were excluded).  The
  Pleiades data are a compilation of the results from \citet{ple1},
  \citet{ple2}, \citet{ple3}, \citet{ple4}, \citet{ple5}, \citet{ple6},
  \citet{ple7} (taken from the open cluster database), \citet{t99} and
  \citet{se2004}.  Symbols for M34 as Figure \ref{m34_pcd}, with the
  greyscales showing the completeness for $0.02\ {\rm mag}$ periodic
  variations from Figure \ref{m34_sim_results}.}

\label{m34_pmd}
\end{figure}

\begin{figure}
\centering
\includegraphics[angle=270,width=3in]{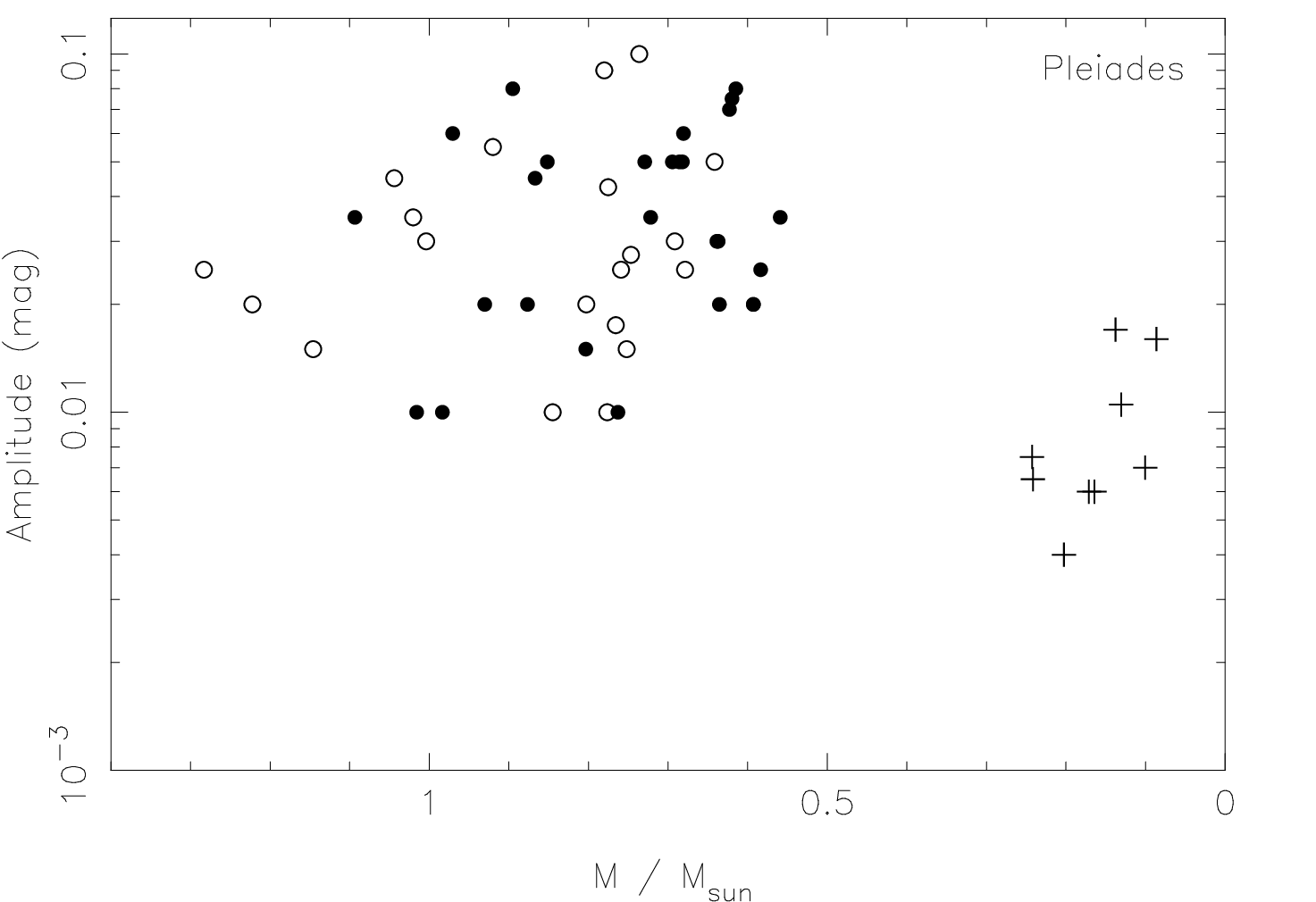}
\includegraphics[angle=270,width=3in]{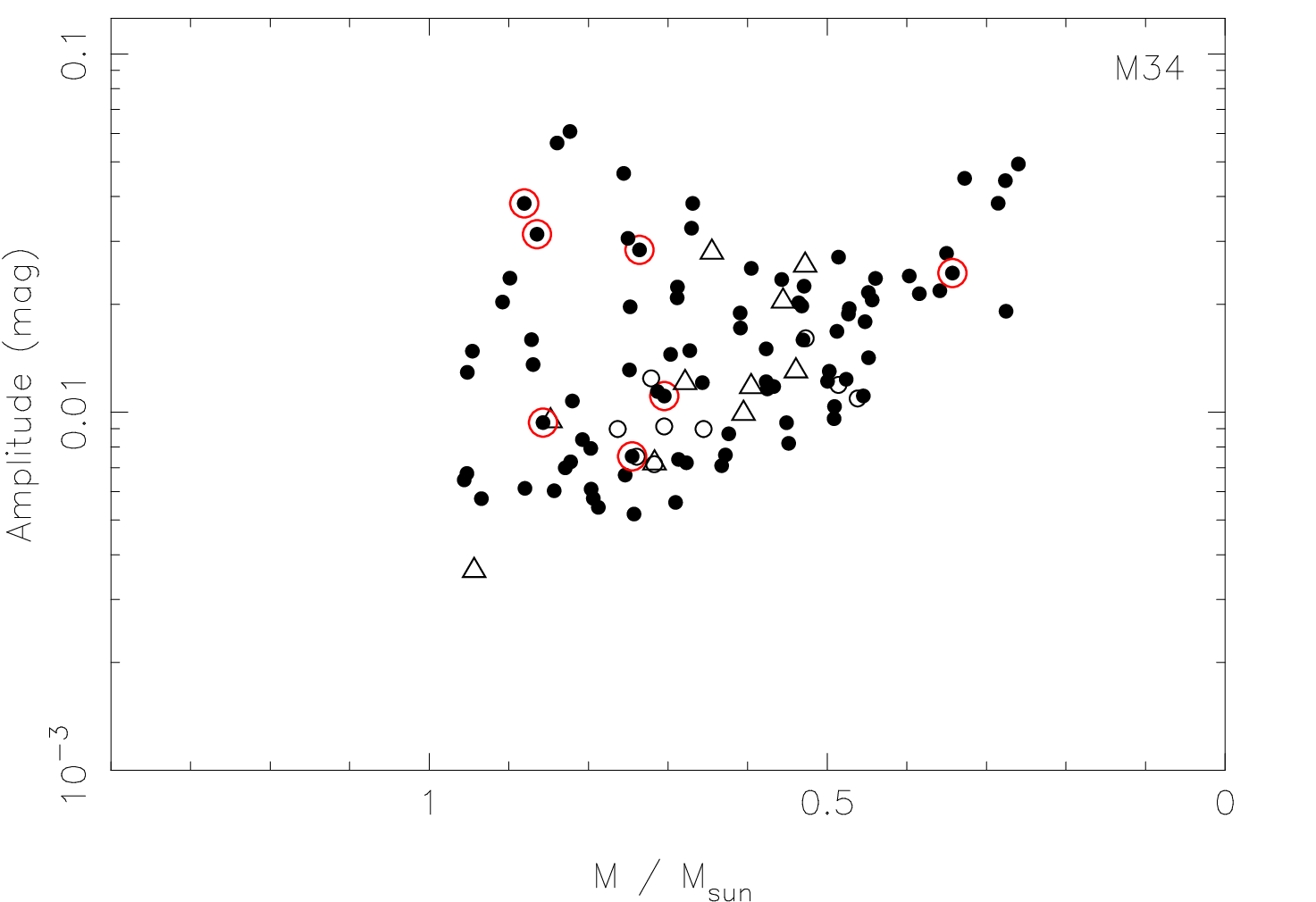}

\caption{Plot of amplitude as a function of mass for the Pleiades
  (top) and M34 (bottom).  The Pleiades amplitudes have been converted
  to the semi-amplitude as used in this work, and the passband used is
  indicated by the symbols: filled circles indicate $V$-band, crosses
  indicate $I$-band, and open circles denote objects with amplitudes
  measured in unknown passbands (but suspected to be $V$).  All
  symbols in the M34 plot are as Figure \ref{m34_pmd} with all objects
  measured in i-band.  The lower envelope of this diagram clearly
  traces the RMS curve in Figure \ref{m34_rms} and therefore results
  from incompleteness at low amplitude for faint stars resulting from
  increased noise.}

\label{m34_amd}
\end{figure}

Period distributions for the objects photometrically selected as
possible cluster members are shown in Figure \ref{m34_perioddist},
a plot of period as a function of $V-I$ colour is shown in Figure
\ref{m34_pcd}, and plots of period and amplitude as a function of mass
are shown in Figures \ref{m34_pmd} and \ref{m34_amd}, respectively.

\subsubsection{Period distributions}

We have attempted to correct the period distribution for cluster
members in Figure \ref{m34_perioddist} for the effects of
incompleteness and (un)reliability using the simulations described in \S
\ref{sim_section}.  The results are shown in the solid histograms in
Figure \ref{m34_perioddist}.  This histogram was generated as follows:
for each bin of the period distribution as a function of {\em
  detected} period, containing $N_i$ objects, we chose $N_i$ random 
objects falling in that bin from the simulations, and generated a new
histogram using the {\em real} periods of the simulated objects.  We
then averaged over $1000$ realisations of this process.  This has the
effect of converting the histogram plotted as a function of {\em
  detected} period to one plotted as a function of {\em real} period.
This was then corrected for incompleteness by dividing by the
completeness fraction from Figure \ref{m34_sim_results}.

We applied two-sided Kolmogorov-Smirnov (K-S) tests to the corrected
distributions in adjacent mass bins to evaluate whether the visible
differences between the distributions were statistically significant.
In particular, there is an apparent clear distinction between the
distribution for $M > 0.4\ \msun$ and $M < 0.4\ \msun$, in the sense
that there are few slow rotators at low-mass.  The results are
summarised in Table \ref{ks_table}.

\begin{table}
\centering
\begin{tabular}{lll}
\hline
Mass range 1            &Mass range 2            &$P({\rm same})$ \\
\hline
$0.7 \le M/\msun < 1.0$ &$0.4 \le M/\msun < 0.7$ &$0.85$ \\
$0.4 \le M/\msun < 0.7$ &$M < 0.4\ \msun$        &$2 \times 10^{-3}$ \\
\hline
\end{tabular}

\caption{Results of two-sided Kolmogorov-Smirnov tests to evaluate the
probabilities that the period distributions in adjacent mass bins are
drawn from the same parent population.}

\label{ks_table}
\end{table}

The K-S test for the first two mass bins is inconsistent with there
being significant differences in the period distribution.  However,
the test for the $0.4 \le M/\msun < 0.7$ and $M < 0.4\ \msun$ bins
indicates that they differ significantly at the $99.8 \%$ confidence
level, ie. the very low-mass stars ($M < 0.4\ \msun$) show a different
period distribution than the higher-mass stars ($M > 0.4\ \msun$).
This effect is also visible in Figure \ref{m34_pmd}, and is discussed
further in \S \ref{low_mass_section}.

\subsubsection{Bimodality?}

The period distributions for $M > 0.4\ \msun$ in Figure
\ref{m34_perioddist} appear to be bimodal when plotted in linear
period (see also Figure \ref{m34_pmd}), but the apparent bimodality is
much less obvious in $\log_{10}$ period (see the lower right panel of
Figure \ref{m34_perioddist}).  Following \citet{st99}, we
attempted to determine whether this bimodality is statistically
significant by performing a one-sample Kolmogorov-Smirnov test against
a model distribution, taken to be a simple function $N(P) = 30(1 - P /
15)$, fit by eye to the upper envelope of the distribution, over $0 <
P < 15\ {\rm days}$.  The result was a $28 \%$ probability that our
period distribution was indeed drawn from this simple model
distribution.  The high probability for this poor model coupled with
the results in $\log_{10}$ period implies that the apparent bimodality
is not significant.

Truly bimodal distributions are typically seen only in much
younger clusters (eg. the ONC, \citealt{h2002}), although there is
evidence that this may also be the case in the Pleiades
(eg. \citealt{sshj93}) and are generally interpreted as resulting from
the presence of a mixture of disc-locked and unlocked, spun-up stars
(see \S \ref{amevol_section}).  Such an interpretation cannot hold at
the age of M34.

Interestingly, the presence of a short-period peak at $P \sim 1-2\
{\rm days}$ and a longer period peak at $P \sim 7\ {\rm days}$ is
consistent with the empirical view on angular momentum evolution of
\citet{b2003}.  Briefly, the short-period peak corresponds to the $C$
or `convective' sequence, and the long-period peak to the $I$ or
`interface' sequence.  In this interpretation, stars on the $C$
sequence have decoupled radiative and convective zones, with angular
momentum loss due to stellar winds coupled to convective magnetic
fields.  On the $I$ sequence, an interface dynamo couples the zones
leading to increased angular momentum loss \citep{b2003}.  Very
low-mass stars (eg. $M < 0.4\ \msun$ in Figure \ref{m34_perioddist})
are fully convective, so they can only lie on the $C$ sequence and
hence fall at short periods.

\subsubsection{Rapid rotators}

The presence of rapid rotation (eg. $P < 1\ {\rm day}$) at the age of
M34 is important for constraining models of angular momentum evolution,
discussed in \S \ref{rotevol_section}.  We have examined the periods
of these stars, to check that they are not rotating close to their
break-up velocity, where they would be short-lived, suggestive of
spin-up on the ZAMS.  The corresponding critical period $P_{crit}$ for
break-up is given approximately by: 
\begin{equation}
P_{\rm crit} = 0.116\ {\rm days}\ {(R / {\rm R_{\odot}})^{3/2}\over{(M / {\rm M_{\odot}})^{1/2}}}
\end{equation}
where $R$ and $M$ are the stellar radius and mass respectively
(eg. \citealt{h2002}).  Using the models of \citet{bcah98},
the object rotating closest to its critical period is M34-1-2402, at
$P = 0.260\ {\rm days}$, where $P_{\rm crit} = 0.039\ {\rm days}$, a
factor of $7$ shorter, so we find no objects in M34 rotating close to
break-up.

\subsubsection{$0.4 - 1.0\ \msun$ objects}

Comparing the distributions of rotational periods for M34 and the
Pleiades in Figure \ref{m34_pmd} indicates that, in general, the M34
distribution resembles a ``spun-down'' Pleiades distribution, exactly
as expected.  However, several conspicuous objects are visible in
Figure \ref{m34_pmd}, with periods close to $1\ {\rm day}$ and high
masses ($\sim 0.8 - 1.0\ \msun$).  These are somewhat difficult to
explain by spin-down of faster Pleiades rotators, given the lack of $<
1\ {\rm day}$ period rotators there for $M \ga 0.9\ \msun$ (see Figure
\ref{m34_pmd}), but the discrepancy could result from $\sim 0.1\
\msun$ errors in the computation of the masses, dominated by errors in
determining the $I$ magnitudes (in particular for the Pleiades data,
due to the conversions required between the photometric systems used
by the various authors).

The lightcurves for these objects were examined carefully, finding
two very significant detections at unusually large amplitude with
associated X-ray sources, M34-1-2953 and M34-3-523.  Detection in
X-rays is suggestive of youth or binarity.  Since these objects are at
high-mass and hence subject to increased field contamination, it is
possible that they may be field objects.  One further object was
detected in X-rays, M34-4-1092, with a $\sim 1 \%$ amplitude, and
interestingly is the only one of these objects lying away from the
cluster sequence, being $\sim 0.5\ {\rm mag}$ brighter.  We therefore
suspect that it is a binary, to be confirmed spectroscopically.  The
remaining objects, M34-4-2697 and M34-4-2833, are two of our lowest
amplitude rotation candidates ($\sim 0.5 \%$) and are marginal
detections.  Our planned spectroscopy of all the candidates should
allow us to weed out any field contaminants, and to measure $v \sin i$
for any very rapid rotators to confirm their nature. 

On the suggestion of the anonymous referee, we note the strong
resemblance between the M34 results in Figure \ref{m34_pmd} and the
Pleiades $v \sin i$ distribution of \citet{sshj93}.  These authors
also found the majority ($80 \%$) of the rotators to have low
velocities, with an upper branch of rapid rotators comprising the
remainder of the sample.

Accounting for the differences in sensitivity between the surveys,
Figure \ref{m34_amd} shows no significant evolution in amplitude from
the Pleiades to M34.  However, we do find an apparent trend of
decreasing amplitude moving to lower masses, down to $\sim 0.4\ \msun$
in M34.  It is not clear whether a similar trend exists in the
Pleiades over this mass range due to the lack of data.

\subsubsection{$< 0.4\ \msun$ objects}
\label{low_mass_section}

Figure \ref{m34_perioddist} shows a clear and statistically
significant lack of slow rotators in the period distribution for 
$M < 0.4\ \msun$.  This appears to be a real effect, and is not
caused, for example, by selection biases in our sample.  This effect
has been observed previously (eg. \citealt{se2004}) at very low mass,
but we were unable to find significant samples of rotation periods in
the literature for the mass range $0.25 < M/\msun < 0.4$ probed by our
sample, so it is complementary to these previous studies.

Tentatively the data in this mass range suggest that the typical
rotation periods of these stars decrease as a function of decreasing
mass, ie. less massive stars are faster rotators.  This is in good
agreement with the results at very low mass ($M \la 0.25\ \msun$) from
\citet{se2004}, who found the rather slow spin-down of VLM stars
to be better fit by models assuming an exponential angular momentum
loss law, giving $P \propto {\rm e}^t$, rather than the \citet{sk72}
$t^{1/2}$ law that applies for higher-mass stars.  They attribute this
difference to the fully convective nature of the low-mass objects,
which cannot support a large scale magnetic dynamo, which gives rise
to the \citet{sk72} angular momentum loss law in solar-mass stars.  
We cannot make a detailed assessment of these conclusions over the
mass range $0.25 < M/\msun < 0.4$, since we could not find a suitable
equivalent large sample of stars with measured rotation periods in a
young open cluster from the literature.  This will be addressed by the
Monitor project, where we have obtained photometric periods for
low-mass stars in several young clusters, including NGC 2516 ($\sim
150\ {\rm Myr}$, \citealt{j2001}), M50 ($\sim 130\ {\rm Myr}$,
\citealt{kal2003}) and NGC 2362 ($\sim 5\ {\rm Myr}$,
\citealt{m2001}), to be presented in a future publication.

\citet{se2004} found significantly lower amplitudes for $M < 0.25\
\msun$ than for the higher-mass ($M \ga 0.5\ \msun$) Pleiades objects
(see Figure \ref{m34_amd}).  We see no evidence for this in the
interval $0.25 < M/\msun < 0.4$, finding instead amplitudes comparable
to those of the stars in the higher-mass bin.  We cannot test the
result in M34 for $M < 0.25\ \msun$ with the present data-set.

\subsubsection{Comparison with $v \sin i$ measurements}

\begin{figure}
\centering
\includegraphics[angle=270,width=3in]{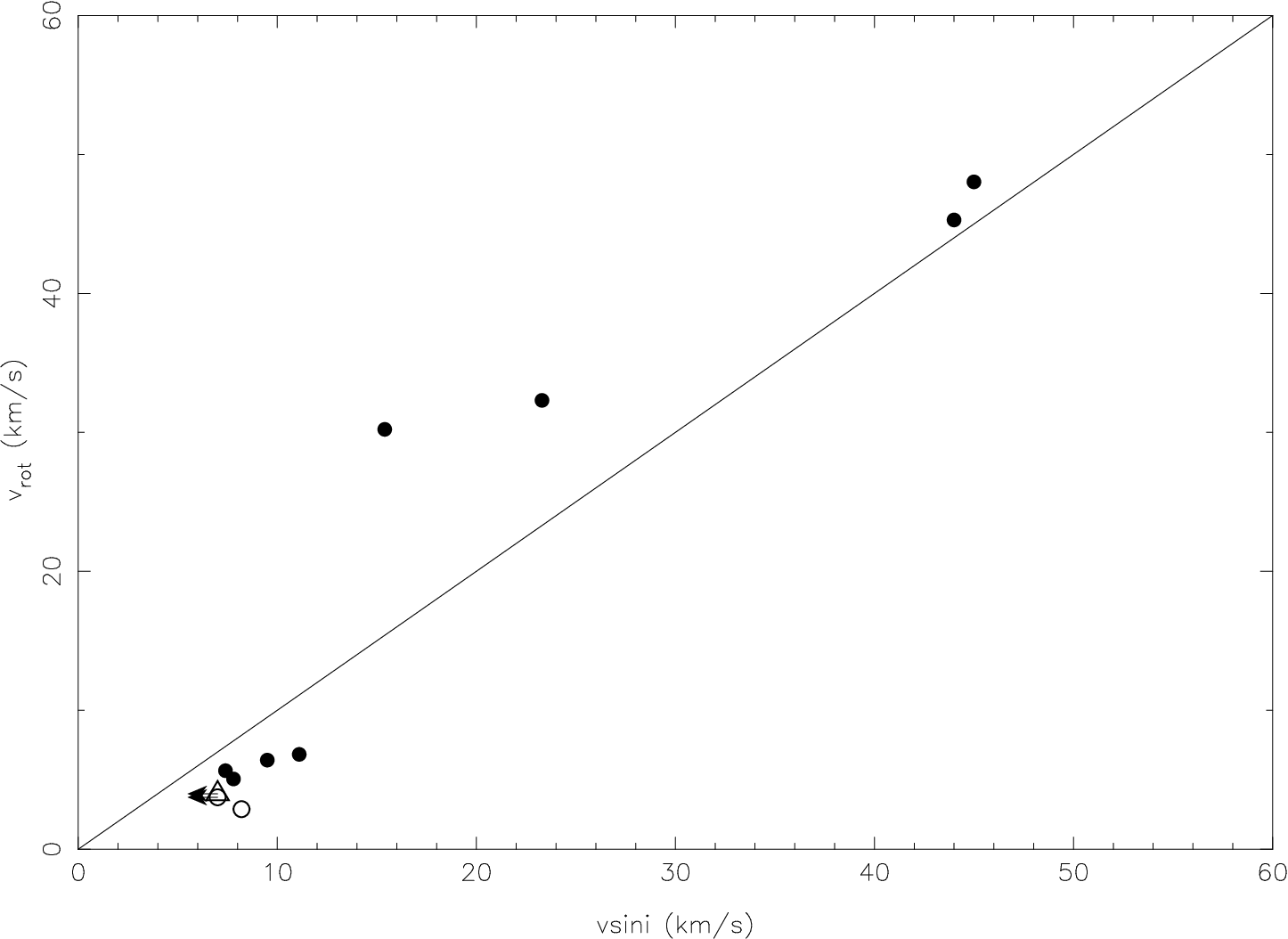}

\caption{Plot of the photometric rotation velocity $v_{\rm rot}$ as a
  function of the $v \sin i$ value measured by \citet{sjf}.  The
  diagonal line indicates the case where $v \sin i = v_{\rm rot}$, and
  the region above the line corresponds to the ``allowed'' region
  where $v \sin i < v_{\rm rot}$.  The small population of objects
  with $v \sin i$ slightly greater than $v_{\rm rot}$ may result from
  over-estimation of the $v \sin i$ values and are close to the lower
  measurable limit for the spectroscopic observations of \citet{sjf}.
  Alternatively, these could highlight deficiencies in the model
  radii we have used to derive $v_{\rm rot}$.}

\label{m34_vsinicomp}
\end{figure}

We have compared our sample of objects with the spectroscopic sample
of \citet{sjf}.  Due to the different magnitude ranges, we only have
$22$ objects in common, and obtained photometric periods for $12$ of
these.  The $10$ objects missed did not show significant periodic
variability during visual examination of the lightcurves.
Figure \ref{m34_vsinicomp} shows a comparison of the $v \sin i$ values
with $v_{\rm rot}$ derived from our photometric periods, derived as:
\begin{equation}
v_{\rm rot} = 2 \pi R / P
\end{equation}
with stellar radii $R$ taken from the models of \citet{bcah98}.

We note that the long-period end of all the figures presented in this
section is somewhat affected by the short observing window we had
available, and should therefore be treated with caution.
Nevertheless, we believe that useful conclusions can still be drawn
from the data.

\subsection{Rotational evolution}
\label{rotevol_section}

\begin{figure*}
\centering
\includegraphics[angle=270,width=6in]{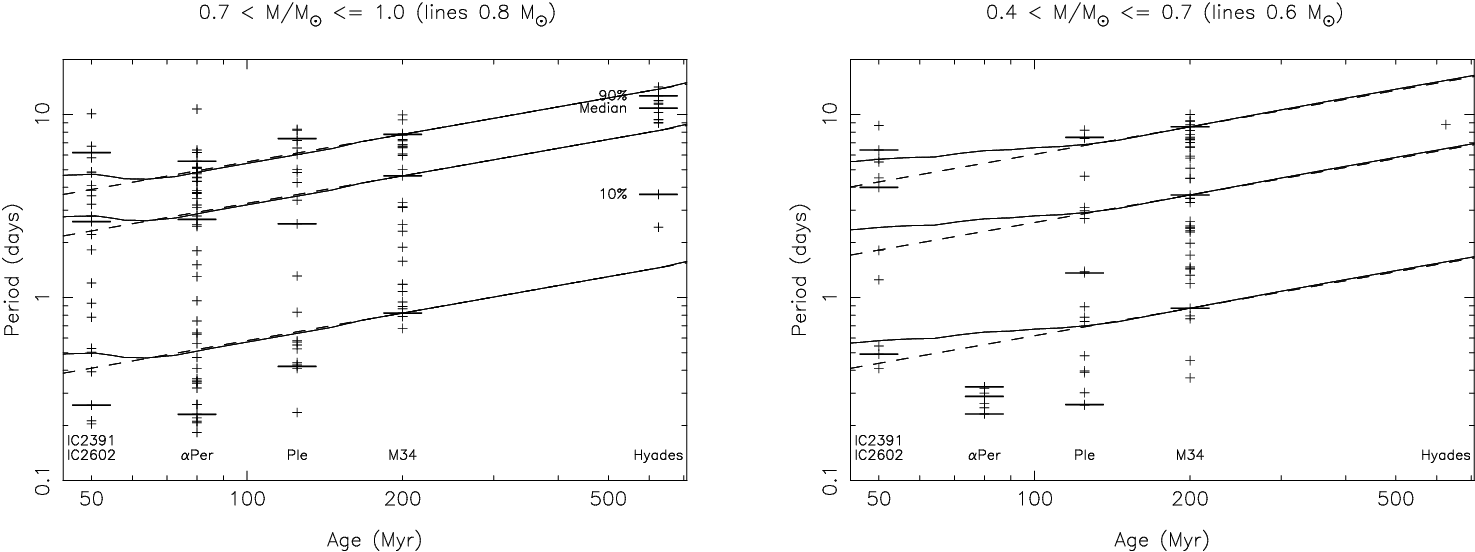}

\caption{Plot of the IC 2391, IC 2602 ($\sim 50\ {\rm Myr}$,
  \citealt{bar2004}), $\alpha$ Per ($\sim 80\ {\rm Myr}$,
  \citealt{s99}), Pleiades, M34 (using all of our period
  detections to avoid biasing the distribution against long periods)
  and Hyades rotation periods as a function of time, in 
  two mass bins:  $0.7 \le M/\msun < 1.0$ 
  (corresponding roughly to G and K spectral types) and $0.4 \le
  M/\msun < 0.7$ (early-M).  A similar diagram was not plotted for the
  $M < 0.4\ \msun$ (late-M and VLM) bin of Figure \ref{m34_perioddist}
  since there were no suitable samples in the other clusters covering
  the same mass range ($0.25 < M/\msun < 0.4$).  The upper and lower
  $10\%$iles of the period distributions, and their medians, 
  are shown as short horizontal lines.  The change in the median
  indicates the convergence toward long periods moving from the
  Pleiades, to M34, to the Hyades.  The IC 2391 data were taken from
  \citet{ps96} and IC 2602 from \citet{bsps99}.  The $\alpha$ Per data
  are a compilation of the results 
  from \citet{aper1}, \citet{aper2}, \citet{ple4}, \citet{aper3},
  \citet{ple5}, \citet{aper4}, \citet{ple6}, \citet{aper5},
  \citet{aper6}, \citet{aper7}, \citet{aper8}, \citet{aper9},
  \citet{aper10}, \citet{aper11}, and the Hyades data from 
  \citet{hya1} and \citet{ple6}, taken from the open cluster
  database.  The solid lines show the predicted evolution from
  Equations \ref{periodevol_equation} and \ref{sk_equation} for a $0.8
  \msun$ star in the left-hand panel, and a $0.6 \msun$ star in the
  right-hand panel.  The dashed lines show only the $t^{1/2}$ angular
  momentum loss component from (\ref{sk_equation}) for comparison.}

\label{periodevol_figure}
\end{figure*}

The ranges of measured rotation periods in IC 2391, IC 2602, $\alpha$
Per, the Pleiades, M34 and the Hyades are compared in Figure
\ref{periodevol_figure}.  The samples for IC 2391 and IC 2602 have
been combined due to the very similar distance moduli and ages of
these clusters.  We use the lower and upper $10\%$iles of the
rotational period distributions to trace the evolution of the rapid
and slow rotators, respectively.  The convergence in rotation rates
across the full age range from IC 2391 and IC 2602 at $\sim 30\ {\rm
  Myr}$ to the Hyades ($\sim 625\ {\rm Myr}$, \citealt{p98}) is
clearly visible in the diagram, due to spin-down of the rapid
rotators by a greater amount than the slow rotators.  The diagram
also indicates the differences in evolution between (roughly) K and
M stars, although it is difficult to constrain the latter because the
samples we have gathered from the literature in $\alpha$ Per and the
Hyades are substantially incomplete at low masses.

We caution that the samples in Figure \ref{periodevol_figure} were
taken from a wide variety of sources and may not be directly
comparable.  The reliability of the long periods in our M34 sample may
be reduced by our limited time base-line.  For the rapid rotators,
the $10\%$ile may be affected by field contamination (due to field
objects misclassified as cluster members), for example contact binary
systems, which have very short photometric periods, but are otherwise
difficult to distinguish from rotational modulation at low amplitude.

The rotational evolution can be simply modelled by assuming that it
results from two components: the spin-up due to contraction in stellar
radius (while conserving total angular momentum) as a function of age,
which we compute from the models of \citet{bcah98}, and spin-down
resulting from angular momentum loss:
\begin{equation}
P_2 = \alpha P_1 \left(R_2 / R_1\right)^2
\label{periodevol_equation}
\end{equation}
where $P_1$ and $P_2$ are the rotation periods at the two ages,
$R_1$ and $R_2$ are the corresponding stellar radii, and $\alpha$ is a
multiplicative factor resulting from angular momentum loss.

The simple \citet{sk72} law gives
\begin{equation}
\alpha_{\rm sk} = \left(t_2 / t_1\right)^{1/2}
\label{sk_equation}
\end{equation}
where $t_1$ and $t_2$ are the respective ages ($t_1 < t_2$).  The
predictions of this model are compared with the observations in Figure
\ref{periodevol_figure}.  The results indicate that in general, the
angular momentum loss law (\ref{sk_equation}) loses angular momentum
too slowly between the Pleiades and M34, and likewise for the fast
rotators between M34 and the Hyades.  It appears to lose angular
momentum (slightly) too quickly for the slow rotators from M34 to the
Hyades, although the latter conclusion depends on our measured slow
rotation periods, which may be unreliable.  We suggest that the
observations require a more rapid spin-down from the Pleiades to M34,
followed by flattening between M34 and the Hyades.

Our results do however appear to be consistent with more comprehensive
models of angular momentum evolution assuming solid body rotation, for
example \citet{bfa97}, where the maximum rotational velocity decreases
from $v_{\rm max} = 100\ {\rm km\ s^{-1}}$ ($P \sim 0.5\ {\rm days}$
at $\msun$, $\rsun$) at the age of the Pleiades to $v_{\rm max} = 50\
{\rm km\ s^{-1}}$ ($P \sim 1\ {\rm day}$) at the age of M34, roughly
as observed.  Models incorporating core-envelope decoupling
(eg. \citealt{a98}, see also Figure 3 of \citealt{b97} and
\citealt{k95}) cause much less efficient braking on the ZAMS for the
first few hundred Myr, giving rise to faster rotators at the age of
M34 than we have detected for these masses.

\section{Conclusions}
\label{conclusions_section}

We have reported on results of a photometric survey of M34 in $V$
and $i$ bands.  Selection of candidate members in a $V$ versus $V-I$
colour-magnitude diagram using an empirical fit to the cluster main
sequence found $714$ candidate members, over a $V$ magnitude range of
$14 < V < 24$ ($0.12 \la M/\msun \la 1.0$).  The likely field
contamination level was estimated using a simulated catalogue of field
objects from the Besan\c{c}on galactic models \citep{r2003},
finding that $\sim 283$ objects were likely field contaminants, an
overall contamination level of $\sim 39 \%$, implying that there are
$\sim 400$ real cluster members over this mass range in our
field-of-view.  The next step in our membership survey will be to
combine our optical data with planned near-IR photometry and optical
spectroscopy to reduce the contamination level.

The M34 mass function was examined over the range $0.1 \la M/\msun \la
1.0$ ($13 < I < 20$), finding that a log-normal function was a good
description of the data in ${\rm dN/dlogM}$, with parameters $M_0
= 0.44 \pm 0.31\ \msun$ (the mass at maximum ${\rm dN/dM}$), and
$\sigma = 0.66 \pm 0.08$.

From $\sim 8\ {\rm nights}$ of time-series photometry (many of which
were partial) we derived lightcurves for $\sim 8500$ objects in the
M34 field, achieving a precision of $< 1 \%$ over $13 \la i \la 17$.
The lightcurves of our candidate M34 cluster members were searched for
periodic variability corresponding to stellar rotation, giving $105$
detections over the mass range $0.25 < M/\msun < 1.0$.

The rotational period distribution for $0.4 < M/\msun < 1.0$ was found
  to peak at $\sim 7\ {\rm days}$ with a tail of fast rotators having
rotational periods down to $\sim 0.8\ {\rm days}$.  Our data suggest
that most of the angular momentum loss for stars with masses $\ga 0.4\
\msun$ happens on the early main sequence, by the age of M34 ($\sim
200\ {\rm Myr}$), in particular for the slow rotators.  We observed
a number of rapidly-rotating stars ($v_{\rm rot} \sim 50-100\ {\rm km\
s^{-1}}$), finding in general that the results can be explained by
models of stellar angular momentum loss assuming solid body rotation
(eg. \citealt{bfa97}) without needing to invoke core-envelope
decoupling, which seems to give rise to faster rotators at the age of
M34 than we have detected for these masses.

Our rotation period distribution for $0.25 < M/\msun < 0.4$ was found
to peak at short periods, with an lack of slow rotators (eg. $P \ga 5\
{\rm days}$).  Our simulations indicate that the effect is real and
does not result from a biased sample.  This is consistent with the
work of other authors (eg. \citealt{se2004}) at very low masses.

The multi-band photometry we have obtained can be used to examine the
properties of the star spots giving rise to the rotational
modulations.  This will be investigated in a future publication.

One of the major problems in our survey is field contamination.  We
intend to publish the final catalogue of M34 membership candidates
after obtaining follow-up spectroscopy.  However, the preliminary
catalogue of membership candidates is available on request.

\section*{Acknowledgments}

The Isaac Newton Telescope is operated on the island of La Palma by
the Isaac Newton Group in the Spanish Observatorio del Roque de los
Muchachos of the Instituto de Astrofisica de Canarias.  This
publication makes use of data products from the Two Micron All Sky
Survey, which is a joint project of the University of Massachusetts
and the Infrared Processing and Analysis Center/California Institute
of Technology, funded by the National Aeronautics and Space
Administration and the National Science Foundation.  This research has
also made use of the SIMBAD database, operated at CDS, Strasbourg,
France.  The Open Cluster Database, as provided by C.F. Prosser
(deceased) and J.R. Stauffer, may currently be accessed at {\tt
  http://www.noao.edu/noao/staff/cprosser/}, or by anonymous ftp to
{\tt 140.252.1.11}, {\tt cd /pub/prosser/clusters/}.

JMI gratefully acknowledges the support of a PPARC studentship, and SA
the support of a PPARC postdoctoral fellowship.  We also thank
Alexander Scholz for his assistance during SA's trip to Toronto, and
the anonymous referee for valuable comments which helped to improve
the paper.

\appendix
\section{Lightcurves and tabular data}
\label{lc_section}

\begin{figure*}
\centering
\includegraphics[angle=0,width=6.4in]{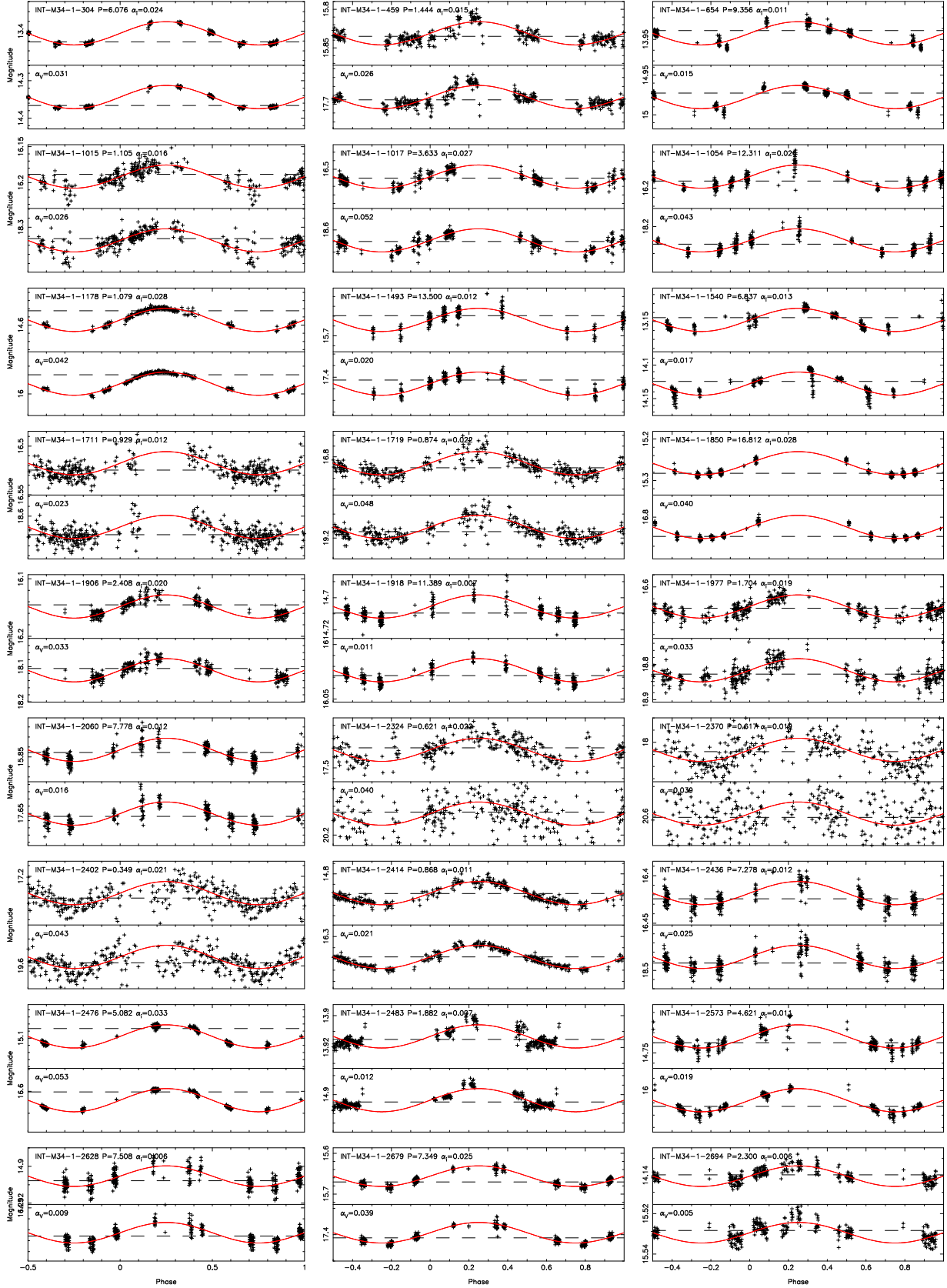}

\caption{Phase-folded lightcurves of our $105$ periodic
  variables classified as candidate cluster members.  The upper
  panel for each object shows the $i$-band lightcurve, the lower panel
  the $V$-band lightcurve.  The dashed lines show the median flux
  level of the object, and the solid lines show the fitted sine
  curves.  The lightcurves are labelled with period $P$ in days, and
  $\alpha_i$ and $\alpha_V$, the amplitudes in $i$ and $V$ bands,
  respectively.  See also Table \ref{m34_cand_table}.}

\label{m34_cand_lc}
\end{figure*}

\begin{figure*}
\centering
\includegraphics[angle=0,width=6.4in]{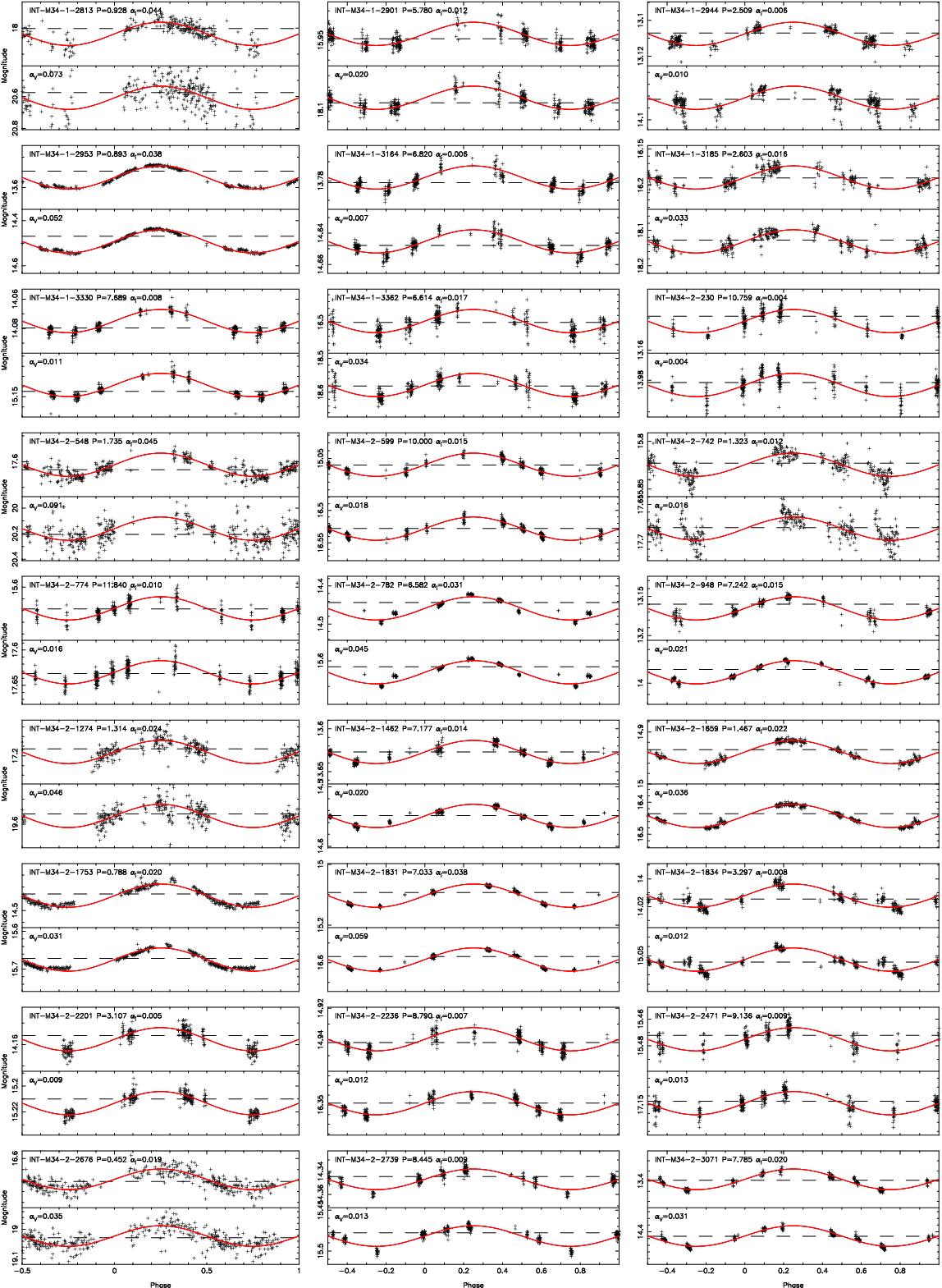}

\contcaption{}

\end{figure*}

\begin{figure*}
\centering
\includegraphics[angle=0,width=6.4in]{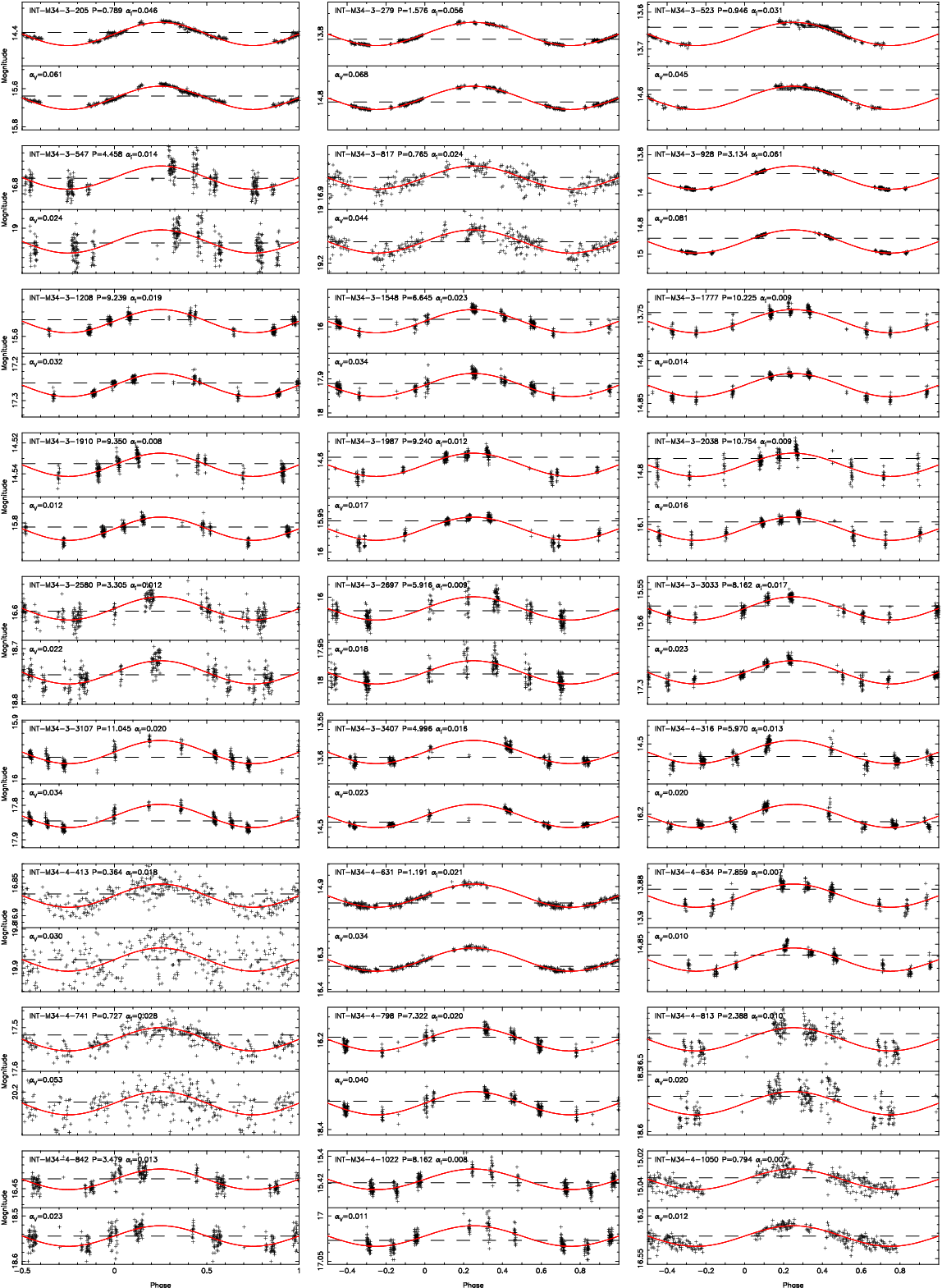}

\contcaption{}

\end{figure*}

\begin{figure*}
\centering
\includegraphics[angle=0,width=6.4in]{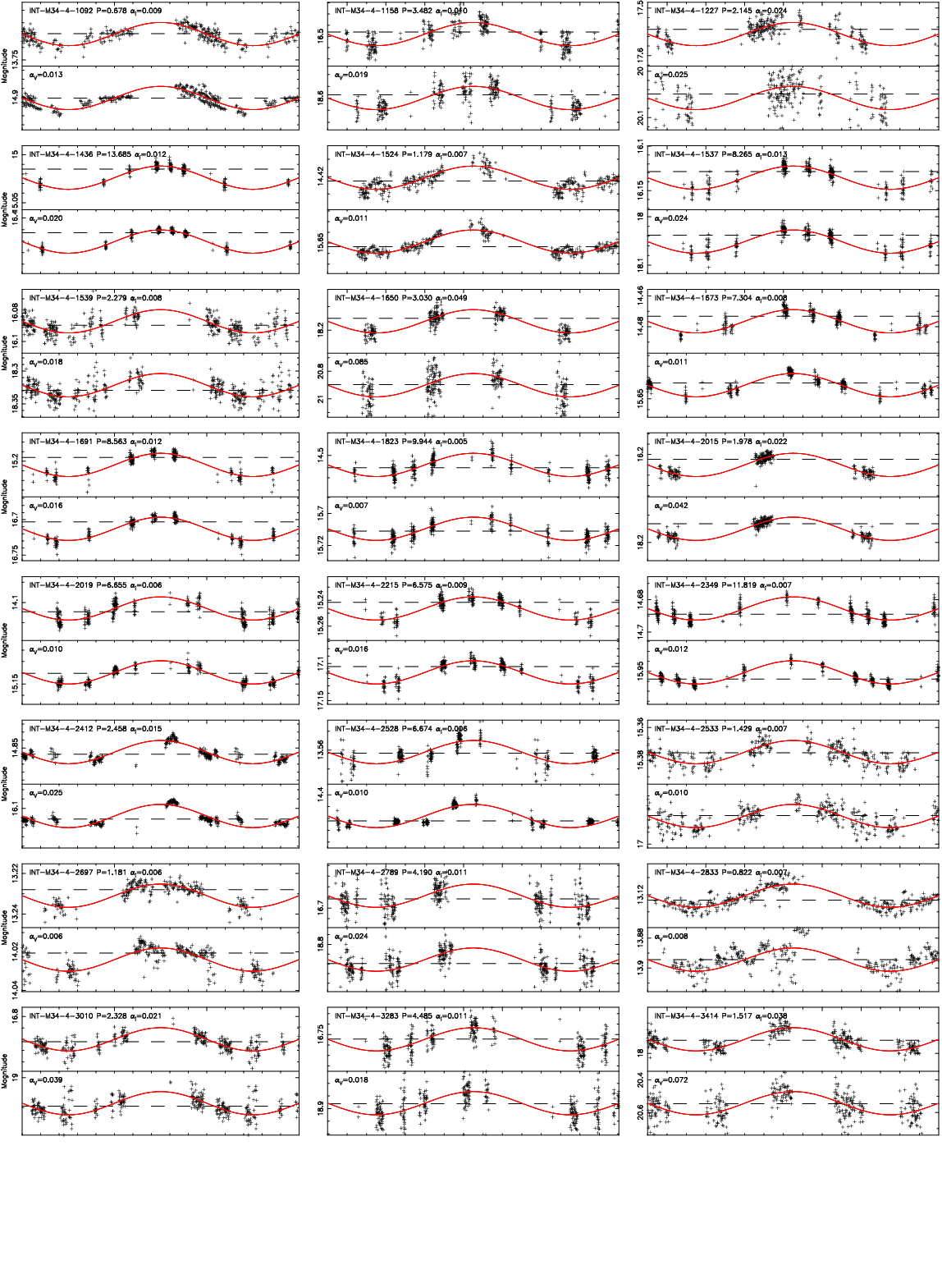}

\contcaption{}

\end{figure*}

\begin{table*}
\centering
\begin{tabular}{lllrrrrrrrrrrr}
\hline
Identifier     &RA    &Dec   &$V$   &$R$   &$I$   &H$\alpha$ &$P$    &$\alpha_V$ &$\alpha_i$ &$M$ &$R$ &JP &$v \sin i$\\
               &J2000 &J2000 &mag   &mag   &mag   &mag       &days   &mag        &mag        &$\msun$ &$\rsun$ & &${\rm km\ s^{-1}}$ \\
\hline
M34-1-304  &02 42 56.88 &+42 35 22.0 &14.52 &13.20 &13.50 &13.91 & 6.076 &0.031 &0.024 &0.90 &0.87 &522 &     \\
M34-1-459  &02 42 51.73 &+42 33 48.9 &17.84 &16.74 &15.79 &16.59 & 1.444 &0.026 &0.015 &0.58 &0.54 &    &     \\
M34-1-654  &02 42 46.41 &+42 39 11.4 &15.12 &13.90 &14.00 &14.44 & 9.356 &0.015 &0.011 &0.82 &0.77 &487 &     \\
M34-1-1015 &02 42 34.26 &+42 31 00.5 &18.47 &17.34 &16.16 &17.14 & 1.105$^{\rm a}$ &0.026 &0.016 &0.53 &0.49 &    &     \\
M34-1-1017 &02 42 35.02 &+42 39 29.0 &18.81 &17.63 &16.46 &17.42 & 3.633 &0.052 &0.027 &0.49 &0.45 &    &     \\
M34-1-1054 &02 42 33.07 &+42 30 02.0 &18.43 &17.28 &16.15 &17.07 &12.311$^{\rm b}$ &0.043 &0.026 &0.53 &0.49 &    &     \\
M34-1-1178 &02 42 31.54 &+42 37 10.5 &16.08 &15.04 &14.58 &15.19 & 1.079 &0.042 &0.028 &0.74 &0.69 &424 & 23.3 \\
M34-1-1493 &02 42 23.32 &+42 38 21.0 &17.56 &16.48 &15.65 &16.41 &13.500$^{\rm b}$ &0.020 &0.012 &0.60 &0.56 &    &     \\
M34-1-1540 &02 42 21.89 &+42 32 13.1 &14.27 &12.83 &13.20 &13.63 & 6.837 &0.017 &0.013 &0.95 &0.92 &377 & 11.1 \\
M34-1-1711 &02 42 16.99 &+42 38 12.0 &18.79 &17.59 &16.45 &17.42 & 0.929$^{\rm a}$ &0.023 &0.012 &0.49 &0.45 &    &     \\
M34-1-1719 &02 42 16.16 &+42 34 01.0 &19.32 &18.03 &16.72 &17.82 & 0.874 &0.048 &0.022 &0.45 &0.41 &    &     \\
M34-1-1850 &02 42 11.67 &+42 31 17.9 &17.01 &15.95 &15.28 &15.92 &16.812$^{\rm b}$ &0.040 &0.028 &0.65 &0.61 &    &     \\
M34-1-1906 &02 42 10.65 &+42 34 09.1 &18.25 &17.11 &16.09 &16.97 & 2.408 &0.033 &0.020 &0.54 &0.50 &    &     \\
M34-1-1918 &02 42 10.30 &+42 34 50.2 &16.18 &15.13 &14.73 &15.28 &11.389$^{\rm b}$ &0.011 &0.007 &0.72 &0.67 &    &     \\
M34-1-1977 &02 42 07.85 &+42 37 45.1 &18.97 &17.77 &16.55 &17.56 & 1.704 &0.033 &0.019 &0.47 &0.43 &    &     \\
M34-1-2060 &02 42 04.92 &+42 37 47.3 &17.80 &16.69 &15.81 &16.60 & 7.778 &0.016 &0.012 &0.58 &0.54 &    &     \\
M34-1-2324 &02 41 55.85 &+42 33 32.5 &20.26 &18.86 &17.32 &18.62 & 0.621 &0.040 &0.022 &0.36 &0.34 &    &     \\
M34-1-2370 &02 41 54.69 &+42 35 29.7 &20.73 &19.41 &17.88 &19.12 & 0.617 &0.039 &0.019 &0.28 &0.28 &    &     \\
M34-1-2402 &02 41 53.31 &+42 32 10.0 &19.70 &18.49 &17.16 &18.23 & 0.349 &0.043 &0.021 &0.38 &0.35 &    &     \\
M34-1-2414 &02 41 53.25 &+42 35 26.3 &16.48 &15.53 &14.82 &15.49 & 0.868 &0.021 &0.011 &0.70 &0.66 &    &     \\
M34-1-2436 &02 41 52.59 &+42 36 01.6 &18.63 &17.47 &16.36 &17.30 & 7.278 &0.025 &0.012 &0.50 &0.46 &    &     \\
M34-1-2476 &02 41 51.70 &+42 38 23.4 &16.75 &15.85 &15.08 &15.75 & 5.082 &0.053 &0.033 &0.67 &0.63 &    &     \\
M34-1-2483 &02 41 51.31 &+42 34 24.8 &15.06 &13.67 &13.98 &14.40 & 1.882 &0.012 &0.007 &0.82 &0.78 &241 &     \\
M34-1-2573 &02 41 48.98 &+42 39 59.7 &16.18 &15.25 &14.75 &15.31 & 4.621 &0.019 &0.011 &0.71 &0.67 &    &     \\
M34-1-2628 &02 41 46.37 &+42 32 31.9 &16.43 &15.50 &14.93 &15.49 & 7.508 &0.009 &0.006 &0.69 &0.65 &    &     \\
M34-1-2679 &02 41 44.20 &+42 35 35.9 &17.56 &16.48 &15.65 &16.39 & 7.349 &0.039 &0.025 &0.60 &0.56 &    &     \\
M34-1-2694 &02 41 43.96 &+42 40 31.7 &15.68 &14.61 &14.15 &14.73 & 2.300 &0.005 &0.006 &0.80 &0.75 &198 &     \\
M34-1-2813 &02 41 39.72 &+42 38 06.7 &20.70 &19.38 &17.88 &19.09 & 0.928 &0.073 &0.044 &0.28 &0.28 &    &     \\
M34-1-2901 &02 41 36.60 &+42 40 03.9 &18.24 &17.02 &15.87 &16.88 & 5.780 &0.020 &0.012 &0.57 &0.53 &    &     \\
M34-1-2944 &02 41 35.08 &+42 33 30.9 &14.23 &12.80 &13.18 &13.59 & 2.509$^{\rm c}$ &0.010 &0.006 &0.96 &0.93 &159 &     \\
M34-1-2953 &02 41 35.25 &+42 41 02.5 &14.63 &13.62 &13.62 &14.05 & 0.893 &0.052 &0.038 &0.88 &0.85 &158 & 45.0 \\
M34-1-3164 &02 41 27.66 &+42 35 42.1 &14.79 &13.46 &13.85 &14.23 & 6.820 &0.007 &0.006 &0.84 &0.80 &131 &     \\
M34-1-3185 &02 41 26.29 &+42 30 14.9 &18.28 &17.13 &16.13 &17.00 & 2.603 &0.033 &0.016 &0.53 &0.49 &    &     \\
M34-1-3330 &02 41 21.40 &+42 35 44.4 &15.29 &14.03 &14.15 &14.59 & 7.689 &0.011 &0.008 &0.80 &0.75 &105 &     \\
M34-1-3362 &02 41 20.03 &+42 39 23.7 &18.74 &17.56 &16.44 &17.38 & 6.614 &0.034 &0.017 &0.49 &0.45 &    &     \\
M34-2-230  &02 40 30.63 &+42 51 01.5 &14.16 &13.24 &13.25 &13.63 &10.759$^{\rm b}$ &0.004 &0.004 &0.94 &0.91 &    &     \\
M34-2-548  &02 40 09.57 &+42 48 37.3 &20.37 &18.99 &17.52 &18.76 & 1.735 &0.091 &0.045 &0.33 &0.32 &    &     \\
M34-2-599  &02 40 49.09 &+42 48 20.4 &16.69 &15.71 &15.06 &15.69 &10.000 &0.018 &0.015 &0.67 &0.63 &    &     \\
M34-2-742  &02 40 15.08 &+42 47 13.6 &17.88 &16.73 &15.79 &16.62 & 1.323 &0.016 &0.012 &0.58 &0.54 &    &     \\
M34-2-774  &02 40 24.27 &+42 46 57.1 &17.82 &16.62 &15.58 &16.52 &11.840$^{\rm b}$ &0.016 &0.010 &0.61 &0.57 &    &     \\
M34-2-782  &02 40 49.66 &+42 46 55.0 &15.79 &15.03 &14.47 &15.00 & 6.582 &0.045 &0.031 &0.75 &0.70 & 18 &     \\
M34-2-948  &02 40 26.58 &+42 45 40.0 &14.16 &13.23 &13.24 &13.61 & 7.242 &0.021 &0.015 &0.95 &0.92 &    &     \\
M34-2-1274 &02 40 38.20 &+42 43 06.6 &19.71 &18.48 &17.08 &18.20 & 1.314 &0.046 &0.024 &0.40 &0.36 &    &     \\
M34-2-1462 &02 40 30.38 &+42 41 51.8 &14.72 &13.61 &13.68 &14.11 & 7.177 &0.020 &0.014 &0.87 &0.83 &    &     \\
M34-2-1659 &02 40 26.71 &+42 40 15.6 &16.62 &15.66 &14.95 &15.60 & 1.467 &0.036 &0.022 &0.69 &0.65 &    &     \\
M34-2-1753 &02 40 48.54 &+42 39 25.7 &15.82 &14.99 &14.49 &15.02 & 0.788 &0.031 &0.020 &0.75 &0.70 &    &     \\
M34-2-1831 &02 40 42.78 &+42 38 59.1 &16.74 &15.74 &15.09 &15.74 & 7.033 &0.059 &0.038 &0.67 &0.63 &    &     \\
M34-2-1834 &02 40 05.98 &+42 38 57.4 &15.22 &14.25 &14.08 &14.51 & 3.297 &0.012 &0.008 &0.81 &0.76 &    &     \\
M34-2-2201 &02 40 30.81 &+42 36 25.0 &15.38 &14.38 &14.21 &14.65 & 3.107 &0.009 &0.005 &0.79 &0.74 &    &     \\
M34-2-2236 &02 40 19.31 &+42 36 12.9 &16.51 &15.56 &14.96 &15.53 & 8.790 &0.012 &0.007 &0.69 &0.65 &    &     \\
M34-2-2471 &02 40 57.86 &+42 34 39.3 &17.30 &16.23 &15.44 &16.17 & 9.136 &0.013 &0.009 &0.62 &0.58 &    &     \\
M34-2-2676 &02 40 53.92 &+42 33 16.5 &19.18 &17.95 &16.54 &17.66 & 0.452 &0.035 &0.019 &0.47 &0.43 &    &     \\
\hline
\end{tabular}

\caption{Summary of the properties of our $105$ M34 rotation
  candidates.  The period $P$ in days, and amplitudes $\alpha_V$ and
  $\alpha_i$ in the $V$ and $i$ bands (units of magnitudes, in the
  instrumental bandpass), interpolated mass and radius (from the
  models of \citealt{bcah98}, derived using the $I$ magnitudes),
  identification number of \citet{jp96}, and $v \sin i$ measurements
  of \citet{sjf} are given (where available).  The following flags are
  used as superscripts to the period: `a' indicating cases where the
  period was ambiguous (with both long and short periods fitting
  equally well to the observations), `b' indicating a
  poorly-constrained long period (due to the short observing window)
  and `c' indicating objects where the $V$-band detection was
  saturated, so the $i$-band only was used to fit the period (NB. the
  $V$-band magnitudes and amplitudes are unreliable for these
  objects).}
\label{m34_cand_table}
\end{table*}

\begin{table*}
\centering
\begin{tabular}{lrrrrrrrrrrrrr}
\hline
Identifier     &RA    &Dec   &$V$   &$R$   &$I$   &H$\alpha$ &$P$    &$\alpha_V$ &$\alpha_i$ &$M$ &$R$ &JP &$v \sin i$\\
               &J2000 &J2000 &mag   &mag   &mag   &mag       &days   &mag        &mag        &$\msun$ &$\rsun$ & &${\rm km\ s^{-1}}$ \\
\hline
M34-2-2739 &02 40 33.29 &+42 32 54.3 &15.64 &14.76 &14.38 &14.84 & 8.445$^{\rm a}$ &0.013 &0.009 &0.76 &0.71 &    &     \\
M34-2-3071 &02 40 48.91 &+42 30 34.1 &14.60 &13.51 &13.45 &13.88 & 7.785 &0.031 &0.020 &0.91 &0.88 &    &     \\
M34-3-205  &02 42 59.94 &+42 58 01.5 &15.81 &14.64 &14.44 &14.94 & 0.789 &0.061 &0.046 &0.76 &0.71 &536 & 44.0 \\
M34-3-279  &02 42 57.82 &+42 58 03.8 &14.97 &13.67 &13.87 &14.25 & 1.576 &0.068 &0.056 &0.84 &0.80 &524 &     \\
M34-3-523  &02 42 50.76 &+42 58 07.8 &14.75 &13.58 &13.71 &14.09 & 0.946 &0.045 &0.031 &0.86 &0.83 &499 &     \\
M34-3-547  &02 42 49.81 &+43 00 35.3 &19.20 &17.97 &16.72 &17.74 & 4.458 &0.024 &0.014 &0.45 &0.41 &    &     \\
M34-3-817  &02 42 39.42 &+42 55 27.7 &19.30 &18.06 &16.78 &17.83 & 0.765 &0.044 &0.024 &0.44 &0.40 &    &     \\
M34-3-928  &02 42 36.30 &+42 54 31.4 &15.07 &13.73 &13.98 &14.41 & 3.134 &0.081 &0.061 &0.82 &0.78 &444 &     \\
M34-3-1208 &02 42 24.90 &+42 53 25.9 &17.43 &16.39 &15.55 &16.25 & 9.239 &0.032 &0.019 &0.61 &0.57 &    &     \\
M34-3-1548 &02 42 10.29 &+42 59 35.9 &18.08 &16.90 &15.94 &16.80 & 6.645 &0.034 &0.023 &0.56 &0.52 &    &     \\
M34-3-1777 &02 42 02.26 &+43 01 13.3 &15.01 &13.45 &13.82 &14.25 &10.225$^{\rm b}$ &0.014 &0.009 &0.85 &0.81 &288 &  7.0 \\
M34-3-1910 &02 41 57.92 &+42 53 22.4 &15.97 &14.71 &14.56 &15.09 & 9.350$^{\rm a}$ &0.012 &0.008 &0.74 &0.69 &268 &  7.0 \\
M34-3-1987 &02 41 55.95 &+42 58 30.8 &16.17 &14.71 &14.69 &15.24 & 9.240$^{\rm a}$ &0.017 &0.012 &0.72 &0.68 &    &     \\
M34-3-2038 &02 41 54.26 &+42 59 35.6 &16.29 &15.01 &14.82 &15.36 &10.754$^{\rm a}$ &0.016 &0.009 &0.70 &0.66 &253 &     \\
M34-3-2580 &02 41 36.18 &+42 54 55.7 &18.91 &17.70 &16.52 &17.51 & 3.305 &0.022 &0.012 &0.48 &0.44 &    &     \\
M34-3-2697 &02 41 32.72 &+43 02 16.3 &18.17 &16.98 &15.98 &16.85 & 5.916 &0.018 &0.009 &0.55 &0.51 &    &     \\
M34-3-3033 &02 41 18.44 &+42 58 21.3 &17.45 &16.34 &15.55 &16.27 & 8.162 &0.023 &0.017 &0.61 &0.57 &    &     \\
M34-3-3107 &02 41 20.45 &+42 58 52.3 &18.02 &16.76 &15.95 &16.75 &11.045$^{\rm b}$ &0.034 &0.020 &0.56 &0.52 &    &     \\
M34-3-3407 &02 41 05.13 &+42 56 43.1 &14.65 &13.46 &13.67 &14.00 & 4.996 &0.023 &0.016 &0.87 &0.84 & 49 &     \\
M34-4-316  &02 42 57.86 &+42 41 46.7 &16.35 &15.22 &14.49 &15.19 & 5.970 &0.020 &0.013 &0.75 &0.70 &527 &     \\
M34-4-413  &02 42 55.14 &+42 50 40.8 &20.01 &18.57 &16.69 &18.24 & 0.364 &0.030 &0.018 &0.45 &0.41 &    &     \\
M34-4-631  &02 42 47.53 &+42 45 46.8 &16.47 &15.52 &14.94 &15.53 & 1.191 &0.034 &0.021 &0.69 &0.65 &    &     \\
M34-4-634  &02 42 47.72 &+42 47 42.9 &15.00 &13.97 &13.94 &14.34 & 7.859 &0.010 &0.007 &0.83 &0.79 &489 &  7.8 \\
M34-4-741  &02 42 43.65 &+42 45 41.8 &20.37 &19.07 &17.38 &18.68 & 0.727 &0.053 &0.028 &0.35 &0.33 &    &     \\
M34-4-798  &02 42 41.81 &+42 46 01.8 &18.44 &17.26 &16.12 &17.10 & 7.322 &0.040 &0.020 &0.53 &0.49 &    &     \\
M34-4-813  &02 42 41.06 &+42 44 22.2 &18.69 &17.53 &16.42 &17.34 & 2.388 &0.020 &0.010 &0.49 &0.45 &    &     \\
M34-4-842  &02 42 40.62 &+42 48 55.6 &18.69 &17.52 &16.37 &17.31 & 3.479 &0.023 &0.013 &0.50 &0.46 &    &     \\
M34-4-1022 &02 42 33.96 &+42 43 26.2 &17.17 &16.08 &15.41 &16.07 & 8.162 &0.011 &0.008 &0.63 &0.59 &    &     \\
M34-4-1050 &02 42 33.61 &+42 49 12.2 &16.66 &      &15.03 &15.66 & 0.794 &0.012 &0.007 &0.68 &0.64 &    &     \\
M34-4-1092 &02 42 32.28 &+42 49 05.9 &15.05 &13.87 &13.76 &14.25 & 0.678 &0.013 &0.009 &0.86 &0.82 &425 & 17.3 \\
M34-4-1158 &02 42 28.95 &+42 42 10.5 &18.74 &17.57 &16.42 &17.37 & 3.482 &0.019 &0.010 &0.49 &0.45 &    &     \\
M34-4-1227 &02 42 26.30 &+42 43 15.9 &20.18 &18.88 &17.42 &18.63 & 2.145 &0.025 &0.024 &0.34 &0.33 &    &     \\
M34-4-1436 &02 42 20.30 &+42 49 05.5 &16.57 &15.61 &15.02 &15.60 &13.685$^{\rm b}$ &0.020 &0.012 &0.68 &0.64 &    &     \\
M34-4-1524 &02 42 17.23 &+42 48 18.6 &15.80 &14.95 &14.45 &14.97 & 1.179 &0.011 &0.007 &0.75 &0.70 &356 & 15.4 \\
M34-4-1537 &02 42 16.19 &+42 43 11.3 &18.17 &17.07 &16.07 &16.92 & 8.265$^{\rm b}$ &0.024 &0.013 &0.54 &0.50 &    &     \\
M34-4-1539 &02 42 17.30 &+42 51 29.8 &18.47 &17.25 &16.00 &17.03 & 2.279 &0.018 &0.008 &0.55 &0.51 &    &     \\
M34-4-1650 &02 42 12.54 &+42 49 28.5 &21.00 &19.62 &18.00 &19.35 & 3.030 &0.085 &0.049 &0.26 &0.27 &    &     \\
M34-4-1673 &02 42 11.71 &+42 43 48.2 &15.77 &14.93 &14.51 &14.98 & 7.304 &0.011 &0.008 &0.75 &0.70 &330 &     \\
M34-4-1691 &02 42 10.99 &+42 43 16.3 &16.84 &15.87 &15.18 &15.82 & 8.563 &0.016 &0.012 &0.66 &0.62 &    &     \\
M34-4-1823 &02 42 07.50 &+42 47 26.7 &15.85 &      &14.53 &15.03 & 9.944 &0.007 &0.005 &0.74 &0.69 &310 &     \\
M34-4-2015 &02 42 01.81 &+42 41 58.9 &18.27 &      &16.14 &16.99 & 1.978 &0.042 &0.022 &0.53 &0.49 &    &     \\
M34-4-2019 &02 42 02.55 &+42 51 51.4 &15.30 &14.41 &14.17 &14.59 & 6.655 &0.010 &0.006 &0.79 &0.74 &289 &  7.4 \\
M34-4-2215 &02 41 55.25 &+42 50 31.7 &17.24 &16.14 &15.20 &16.02 & 6.575$^{\rm a}$ &0.016 &0.009 &0.66 &0.62 &    &     \\
M34-4-2349 &02 41 50.31 &+42 44 37.6 &16.09 &      &14.72 &15.22 &11.819$^{\rm a}$ &0.012 &0.007 &0.72 &0.67 &229 &  8.2 \\
M34-4-2412 &02 41 48.48 &+42 49 33.6 &16.27 &15.41 &14.88 &15.43 & 2.458 &0.025 &0.015 &0.70 &0.66 &218 &     \\
M34-4-2528 &02 41 44.17 &+42 46 07.5 &14.56 &      &13.62 &13.97 & 6.674 &0.010 &0.006 &0.88 &0.85 &199 &  9.5 \\
M34-4-2533 &02 41 43.86 &+42 45 08.1 &17.10 &16.13 &15.37 &16.03 & 1.429 &0.010 &0.007 &0.63 &0.59 &    &     \\
M34-4-2697 &02 41 38.14 &+42 44 04.4 &14.15 &13.30 &13.30 &13.64 & 1.181$^{\rm c}$ &0.006 &0.006 &0.93 &0.90 &177 &     \\
M34-4-2789 &02 41 34.88 &+42 48 52.7 &18.98 &17.75 &16.62 &17.61 & 4.190$^{\rm a}$ &0.024 &0.011 &0.46 &0.42 &    &     \\
M34-4-2833 &02 41 33.43 &+42 42 11.8 &14.04 &      &13.20 &13.54 & 0.822$^{\rm c}$ &0.008 &0.007 &0.95 &0.92 &148 &     \\
M34-4-3010 &02 41 26.73 &+42 51 34.1 &19.24 &18.04 &16.75 &17.81 & 2.328 &0.039 &0.021 &0.44 &0.40 &    &     \\
M34-4-3283 &02 41 16.53 &+42 49 34.8 &19.04 &17.85 &16.67 &17.66 & 4.485 &0.018 &0.011 &0.45 &0.41 &    &     \\
M34-4-3414 &02 41 11.55 &+42 46 22.4 &20.66 &19.37 &17.81 &19.08 & 1.517 &0.072 &0.038 &0.29 &0.29 &    &     \\
\hline
\end{tabular}

\contcaption{}

\end{table*}

\end{document}